\documentclass[aps,tightenlines,showpacs,nofootinbib,superscriptaddress]{revtex4-1}
\usepackage{latexsym}
\usepackage[dvips]{graphicx}
\usepackage{color}
\usepackage{enumerate}
\newcommand{\beq}{\begin{eqnarray}}
\newcommand{\eeq}{\end{eqnarray}}
\newcommand{\eq}{eqnarray}

\newcommand{\al}{{\alpha}}
\newcommand{\be}{{\beta}}
\newcommand{\ci}{\cite}

\newcommand{\ep}{{\epsilon}}

\newcommand{\De}{\Delta}

\newcommand{\la}{{\lambda}}
\newcommand{\La}{{\Lambda}}

\newcommand{\om}{{\omega}}
\newcommand{\Om}{{\Omega}}

\newcommand{\no}{{\nonumber}}
\newcommand{\f}{\frac}
\newcommand{\ra}{\rightarrow}

\newcommand{\Sch}{Schwarzschild }

\begin{document}

\preprint{arXiv:1508.xxxx [hep-th]}
%\today\\

\title{Complete Classification of Four-Dimensional Black Hole and Membrane Solutions in IR-modified Ho\v{r}ava Gravity}

\author{Carlos Arg\"uelles \footnote{E-mail address: carlos.arguelles@icranet.org}}
\affiliation{ICRANet, P.zza della Repubblica 10, I--65122 Pescara, Italy}
\affiliation{Dipartimento di Fisica, Sapienza Universit\`a di Roma, P.le Aldo Moro 5, I--00185 Rome, Italy}

\author{Nicol\'as Grandi \footnote{E-mail address: grandi@fisica.unlp.edu.ar}}
\affiliation{Instituto de F\'\i sica de La Plata - CONICET \& Dto. de F\'isica - Universidad Nacional de La Plata,\\ C.C. 67, 1900 La Plata, Argentina}
\affiliation{Abdus Salam International Center for Theoretical Physics, Associate Scheme \\ Strada Costiera 11, 34151, Trieste, Italy}

\author{Mu-In Park \footnote{E-mail address: muinpark@gmail.com,~Corresponding author}}
\affiliation{Research Institute for Basic Science, Sogang University,\\ Seoul, 121-742, Korea \\~~}

\begin{abstract}
Ho\v{r}ava gravity has been proposed as a renormalizable, higher-derivative gravity without ghost problems, by considering different scaling dimensions for space and time. In the non-relativistic higher-derivative generalization of Einstein gravity, the meaning and physical properties of black hole and membrane space-times are quite different from the conventional ones. Here, we study the singularity and horizon structures of such geometries in IR-modified Ho\v{r}ava gravity, where the so-called ``detailed balance" condition is softly broken in IR. We classify all the viable static solutions without naked singularities and study its close connection to non-singular cosmology solutions. We find that, in addition to the usual point-like singularity at $r=0$, there exists a ``surface-like" curvature singularity at finite $r=r_S$ which is the cutting edge of the real-valued space-time. The degree of divergence of such singularities is milder than those of general relativity, and the Hawking temperature of the horizons diverges when they coincide with the singularities. As a byproduct we find that, in addition to the usual
``asymptotic limit", a consistent flow of coupling constants, that we called
``GR flow limit", is needed in order to recover general relativity in the IR.
\end{abstract}

\pacs{04.20.Jb, 04.20.Dw, 11.27.+d, 04.60.-m, 04.70.Dy}

\maketitle

\newpage

\section{Introduction}

In 2009, Ho\v{r}ava proposed a renormalizable gravity theory with improved
ultraviolet (UV) behavior, which reduces to Einstein gravity with a non-vanishing cosmological constant in infrared (IR). Such improved behavior is obtained at the price of abandoning Einstein's equal-footing treatment of space and time \ci{Hora:08,Hora}. Since then, various aspects of the theory and its solutions have been studied \ci{Calc,Taka,Kiri,Klus,Lu,Muko:0904,Bran,Cai,Cai:0904,Piao,Gao,Colg,Myun,Orla,Keha,Ghod,
Nast,Soti,Muko:0905_1,Nish,Chen:0905_1,Chen:0905_2,Muko:0905_2,Kono,Char,Li,Kim,
Sari,Calc:0905,Park:0905,Bott:0906,Bellorin:2014qca, Bellorin:2015oja,Park:0906,Kiri:0910,Cai:0910,Argu:1008,Cai:1001}.
The original Ho\v{r}ava model satisfying the so-called ``detailed balance'' condition was shown to have several problems \ci{Lu}: (i) A fine-tuning dynamical mechanism is needed, in order to subtract the infinite cosmological constant arising due to the flow of the theory in the IR limit, (ii) the black hole solution in the Ho\v{r}ava model does not recover the usual Schwarzschild-AdS black hole, (iii) for vanishing cosmological constant, the Newtonian potential cannot be obtained in the weak field approximation.\footnote{We are considering only the ``non-projectable" case where there is the space dependance in the lapse function $N$, as well as some possible time dependence. For the study of the Newtonian potential in the ``projectable" case (or its a variant, called ``covariant Ho\v{r}ava-Lifshitz gravity"), where there is only time dependence in $N$, see e.g. \ci{Hora:1007,Muko:1007,Gumr:1109,Lin:1206}.}

In \ci{Keha} an IR modification which contains the flat Minkowski vacuum solution has been studied, by introducing a term proportional to the Ricci scalar of the spatial geometry $\mu ^4 R^{(3)}$. This was called a ``soft-breaking" of the detailed balance condition, with three-dimensional Newton's constant $\sim \mu^{-1}$ \ci{Hora} in the vanishing cosmological constant case.  Later this was generalized to the case with an arbitrary cosmological constant such that the solutions of \ci{Lu} and \ci{Keha} are recovered as some particular limits by introducing the IR-modification term $\omega R^{(3)}$ with a new parameter $\omega$ \ci{Park:0905}. Actually, it turns out that this ``IR-modified Ho\v{r}ava gravity'' does not have the above-mentioned drawbacks of the original Ho\v{r}ava model \ci{Park:0906}.

Recently, the black ``plane" solution \ci{Argu:1008}, and more generally the ``topological" black holes with arbitrary, constant curvature, horizons \ci{Cai,Ghod}, which includes the black hole solutions as the spherical case as well as the hyperbolic and plane membrane solutions, have been studied in the original Ho\v{r}ava model with detailed balance in four dimensions. In this paper, we consider the generalized model with the IR-modification term proportional to $\omega R^{(3)}$ with an arbitrary IR-modification parameter $\omega$. The resulting equations may provide the black membrane geometry without introducing matter, due to the higher spatial-derivative terms which were absent in general relativity. Here, we study the singularity and horizon
structure of such space-times in IR-modified Ho\v{r}ava gravity and classify
all the viable solutions without naked singularities. In particular, we find
that there exists a surface-like curvature singularity at $r=r_S$ as a
cutting edge of our space-time, where the real-valued space-time ends and
unconventional complex-valued metric starts, as well as the usual point-like
singularity at $r=0$ (for some earlier work, see \ci{Cai:1001}). We find that
their degrees of divergence are milder than those of general relativity (GR),
and Hawking temperatures for the black hole and membrane geometries are
finite unless the singularities coincide with the outermost horizons.
And also, we find that the asymptotic limit is not enough to recover the conventional results of GR but we need another limit, called the ``GR flow limit", which achieves a peculiar form of flows of coupling constants.

The plan of this paper is as follows. In Sec. II, we revisit the static black
hole and membrane solutions in four-dimensional GR and we classify all the
viable solutions without naked curvature singularities, in a manner which is
in parallel with the {\it reduced action} approach to IR-modified Ho\v{r}ava
gravity to be pursued later in Sec. III. In Sec. IV, we study the thermodynamics of the black hole and membrane geometries, and find that the Hawking temperature becomes infinity when the curvature singularity sits on the outermost horizon. In Sec. V, we study its close connection to the conditions for the non-singular Friedman-Lema\^\i tre-Robertson-Walker (FLRW) type cosmology. In Sec. VI, we conclude with several remarks.

\section{The D=4 Black Hole and Membrane Solutions in General Relativity }

It is known that in four-dimensional GR with Minkowski vacuum, {\em i.e.,} with vanishing cosmological constant $\Lambda=0$, the black membrane solution has a naked singularity. This situation changes when a cosmological constant is introduced, and in particular for the case of AdS vacuum ($\Lambda <0$) a horizon which hides the singularity appears. This is basically due to the additional ``attraction" caused by the negative cosmological constant, in contrast to ``null" or ``repulsion" for the cases  $\Lambda\geq0$. \footnote{This may also explain why one can have a black hole solution in three-dimensional AdS space, known as BTZ solution, but not in flat or dS space \ci{Bana:9204,Mann:0812,Park:0811}.  }

In the present section, we summarize these known results \ci{Lemo:9404} \footnote{For a more recent, extensive study, see \ci{Lee:1108}} in the context of topological black holes, which describes the black hyperbolic membrane and plane solutions as well as the black hole solution in a unified way \ci{Amin:9604,Mann:9607,Cai:9609,Vanz:9705,Bril:9705}, in parallel with the approach to IR-modified Ho\v{r}ava gravity followed in the next section.

\subsection{ Metric Ansatz and General Solution}

We start by considering the Einstein gravity action with a cosmological
constant $\Lambda$ which reads ($c \equiv 1$)
\begin{\eq}
S_{\rm EH}=\frac{1}{16 \pi G} \int d^4 x \sqrt{-g} \left(R -2 \Lambda \right).
\label{GR_action}
\end{\eq}
We will be interested in static solutions to the above action with a maximally symmetric ({\it i.e.}, constant curvature) two-dimensional slice.
Then, let us consider the following metric ansatz,
\begin{\eq}
  ds^2=-N^2(r) %c^2
  dt^2+\frac{dr^2}{f(r)}+r^2 d \Omega_k\, ,
  \label{ansatz}
\end{\eq}
where the sub-metric
\begin{\eq}
d \Omega_k=\left(\f{d\rho^2}{1-k \rho^2}+\rho^2 d\phi^2\right) \, ,
\label{sub_metric}
\end{\eq}
describes the two-dimensional surface with a constant scalar curvature,
$R^{(2)}=2k$. Without loss of generality, one may take $k=+1,0,-1$ for
spherical, plane, and hyperbolic geometries, respectively. For $k=\pm1$,
this can be written as the standard form in the coordinates $(\theta, \phi)$
\begin{\eq}
d \Omega_k&=&
\left\{
\begin{array}{lll}
d \theta ^2 +\mbox{sin}^2\theta\, d \phi^2,&~~~~~~&(k=+1)\, , \\
d \theta ^2 +\mbox{sinh}^2\theta \,d \phi^2,& ~~~~~~&(k=-1)\, ,
\end{array}
\right.
\end{\eq}
by considering $\rho=\mbox{sin} \theta,\,
\mbox{sinh} \theta$, respectively.

By substituting the metric ansatz into the action (\ref{GR_action}),
the resulting reduced action, after angular integration, is given by
\begin{\eq}
S_{\rm EH}=\frac{\Om_k}{16 \pi G} \int dt dr \frac{N}{\sqrt{f}}\left(2 (k-f-rf') - 2 \La r^2 \right)\ ,
\end{\eq}
where the prime $(')$ denotes the derivative with respect to $r$ and $\Om_k$ is the volume of the two-dimensional surface with curvature $2k$.
The resulting equations of motions read
\begin{\eq}
- (k-f-r f')+ \La r^2 =0 \,,\qquad\qquad\left(\frac{N}{\sqrt{f}}\right)'=0\, ,
\end{\eq}
obtained by varying the functions $N$ and $f$, respectively. One can obtain the general solution as
 \begin{\eq}
N^2=f=k-\f{\La}{3} r^2-\f{2 M}{r}\, ,
\label{GR_solution}
\end{\eq}
by setting $N/\sqrt{f}=\mbox{constant}\equiv 1$ at the spatial infinity, $r=\infty$.
Here $M$ is an integration constant, which agrees with ADM mass for the black hole $(k=+1)$ case, and generally `$4 \pi \times$ ADM mass density' for the flat ($k=0$) and hyperbolic ($k=-1$) membranes.

\subsection{Singularities and Horizons}

In order to make the singularities of the solution explicit, we consider the curvature invariants,
\begin{\eq}
R&=&4 \Lambda\, , \no \\
R^{\mu \nu \al \be}R_{\mu \nu \al \be} &=&\f{8}{3} \Lambda^2 + \f{48 M^2}{r^6}\, ,
\label{singularity:GR}
\end{\eq}
the later manifesting a curvature singularity with the power of $r^{-3}$ at $r=0$, without any $k$ dependence. This singularity needs to be hidden in
our observable space-time, by forming an event horizon around, following the cosmic censorship conjecture \ci{Penr}. Notice that, due to the singularity at $r=0$, we can consider the ranges of $r>0$ and $r<0$ as representing different solutions. Moreover, since the solution for $r<0$ can be mapped into that for $r>0$ by replacing $M \ra -M$, we can restrict our attention to the solution for $r>0$ and consider both signs of the mass, without any loss of generality.

In order to see the horizon structure of the solution, we need to know the positive roots of the cubic polynomial obtained by multiplying $f(r)$ by $-3r$, namely ${\La} r^3-3kr+{6 M}$, whose number can be also obtained by Descartes' rule of signs, as equal to the number of sign changes between consecutive nonzero coefficients, or less than it by an even number.

\begin{figure}
\includegraphics[width=5.4cm,keepaspectratio]{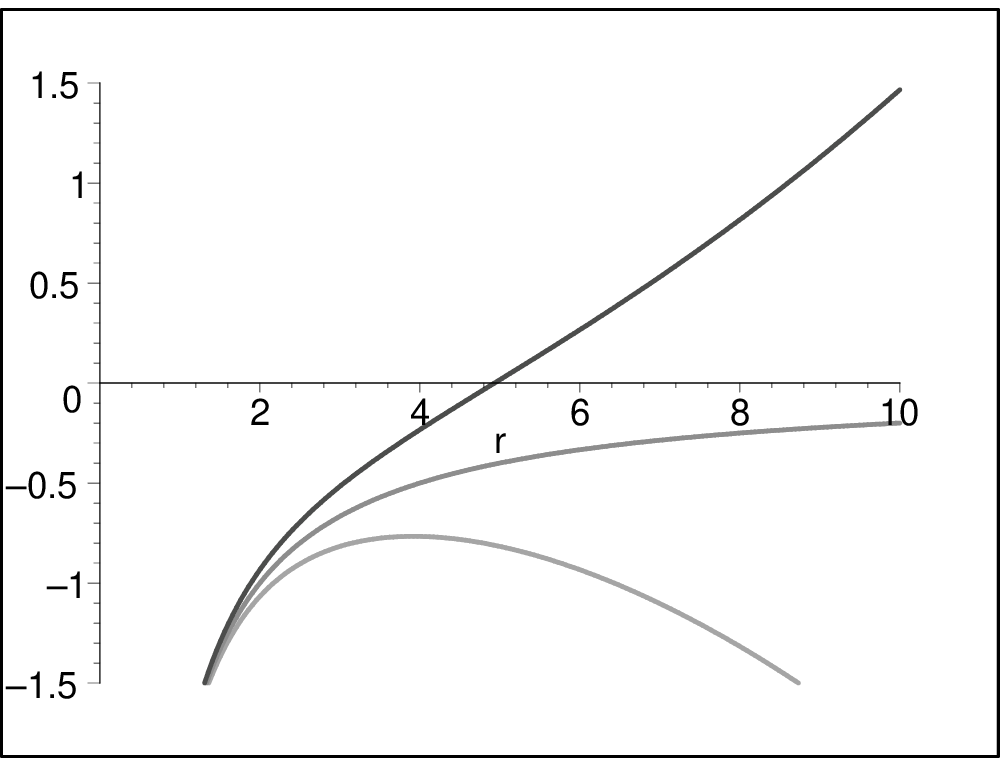}
%{k_zero_M_positive01.eps}
\includegraphics[width=5.4cm,keepaspectratio]{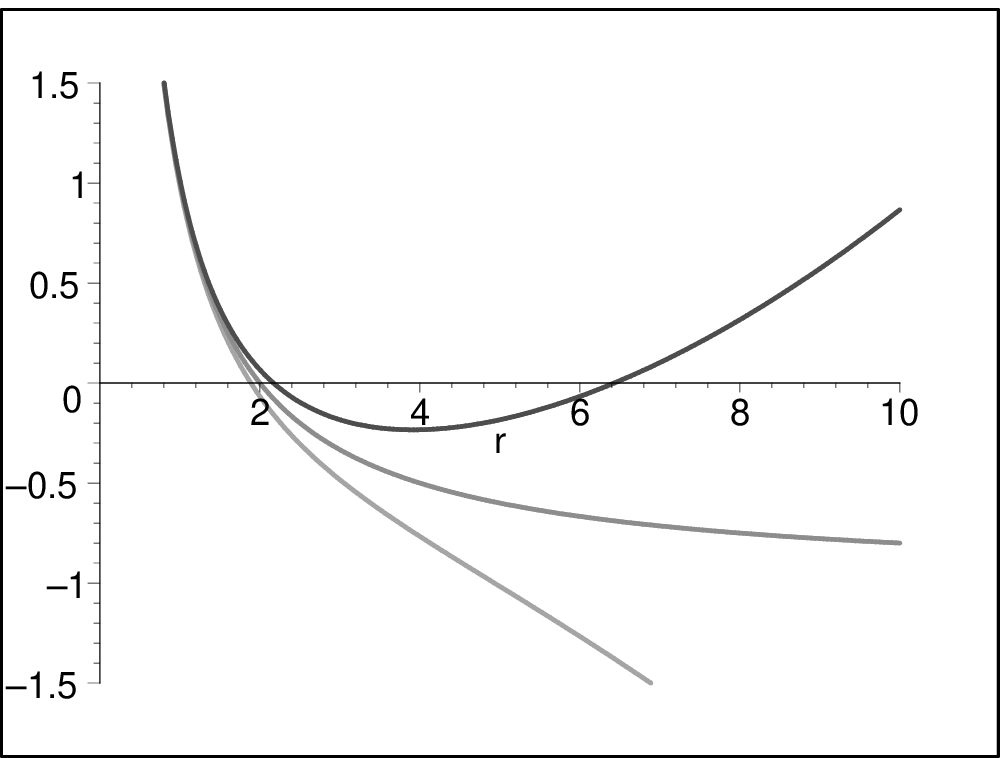}
%{k_negative_M_negative01.eps}
\includegraphics[width=5.4cm,keepaspectratio]{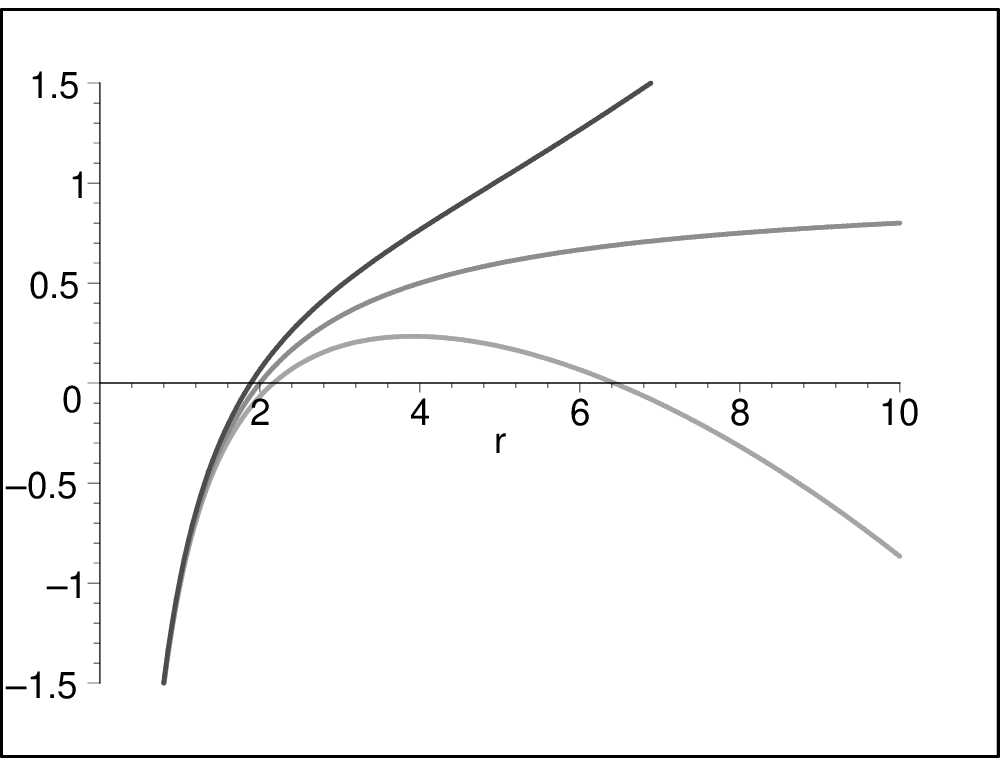}
%{k_positive_M_positive01.eps}
\caption{Plots of $f(r)$ for $k=0~ (M>0)$, $k=-1~ (M<0)$, and $k=+1~(M>0)$
from left to right. For each figure with a given $k$, the three curves
denote AdS ($\Lambda<0$), flat ($\Lambda=0$), dS ($\Lambda>0$) spaces
from top to bottom (we have plotted the AdS/dS cases for $\La=\pm 0.05,
|M|=1$). Due to the curvature singularity at $r=0$, in the case $k=0, M>0$
(left), only the AdS asymptotics (top curve) can be viable due to the
existence of a horizon. The same is true for the AdS asymptotics (top curve)
in the case $k=-1,M<0$ (center), where there are two horizons implying a viable black (hyperbolic) membrane solution without naked singularity. Finally, for  the black hole case $k=+1,M>0$ (right), the AdS (top curve), flat (middle curve), and dS (bottom curve) solutions are all viable. All the remaining curves have a naked singularity at $r=0$ and/or no static region.}
\label{fig:all}
\end{figure}

Let us first consider the {case} $k=0$, {\it i.e.,} the flat membrane
(Fig. \ref{fig:all} (left)). In this case, there is a horizon only if
$\Lambda$ and $M$ have different sign. It is located at
\begin{\eq}
r_+=\left( \f{-6 M}{\Lambda} \right)^{1/3}\, ,
\label{horizon_GR_flat}
\end{\eq}
implying that a ``sensible'' or ``viable'' membrane solution, {\em i.e.,} one in which the singularity is hidden behind a horizon at $r_+$, exists for $M \geq 0$ when $\Lambda <0$ (AdS space). The horizon at $r=r_+$ hides the singularity at $r=0$ and divides the causally connected region of $N^2(r)=f(r)>0$ outside
the horizon (in which $r$ is a space-like variable) from the region of
$N^2(r)=f(r)<0$ inside the horizon (in which $r$ is a time-like variable),
which allows us to interpret
the corresponding solution as a black plane. For $M>0$, we see that
$r_{+} \to \infty$ when $\Lambda \to 0_-$, implying that for the
$\Lambda = 0$ case (flat space) there is neither a horizon at finite
$r$, nor a region in which the coordinate $r$ is space-like, so that this
cannot be considered as a sensible solution. The case $\Lambda>0$ has no
horizon neither, and then again it is not a sensible solution. On the other
hand, for $M <0$, there is a horizon at $r=r_+$, but we will not call this
a ``sensible'' solution since the singularity at $r=0$ is naked as seen
from the region $r<r_+$, where $N^2(r)=f(r)>0$. Of course this solution could
also be interpreted as a time-dependent cosmological solution with $r$ as the
time coordinate, due to the ``equal-footing" treatment of space and time in
GR, but such interpretation will not be possible for Ho\v{r}ava gravity in
the forthcoming sections.
\footnote{Actually, this case corresponds to flipping  $f(r) \ra -f(r)$
together with $(\La,M) \ra (-\La,-M)$ in Fig. \ref{fig:all} (left).}
\begin{figure}
\includegraphics[width=7.3cm,keepaspectratio]{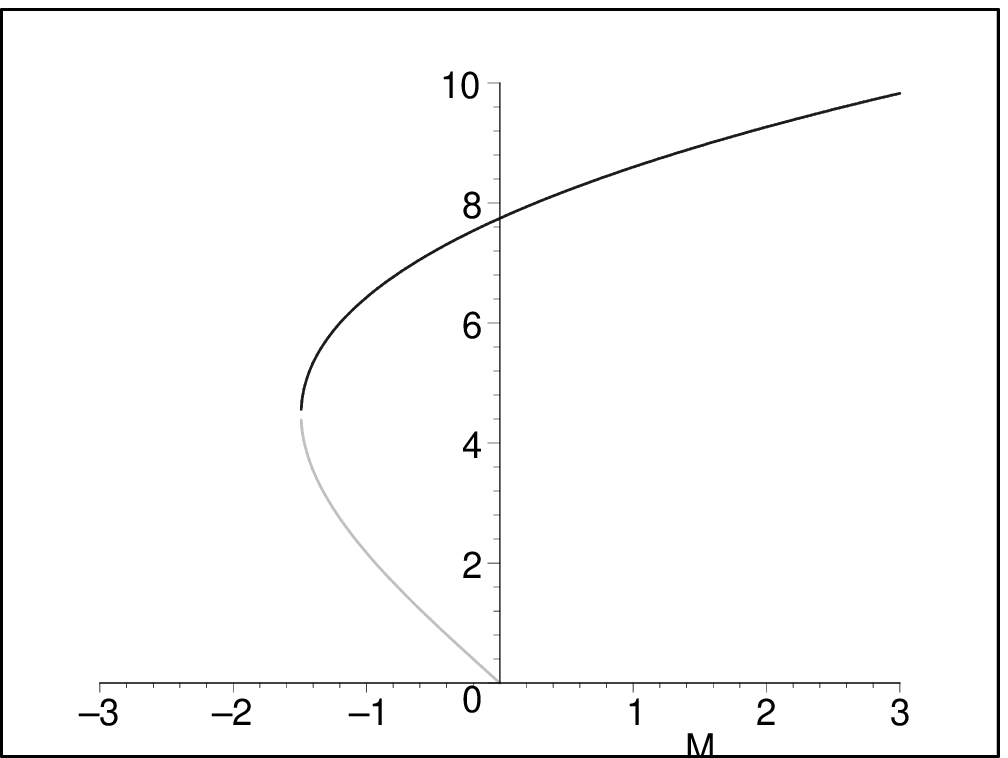}
%{Horizon_k_negative_GR01.eps}
\qquad \qquad
\includegraphics[width=7.3cm,keepaspectratio]{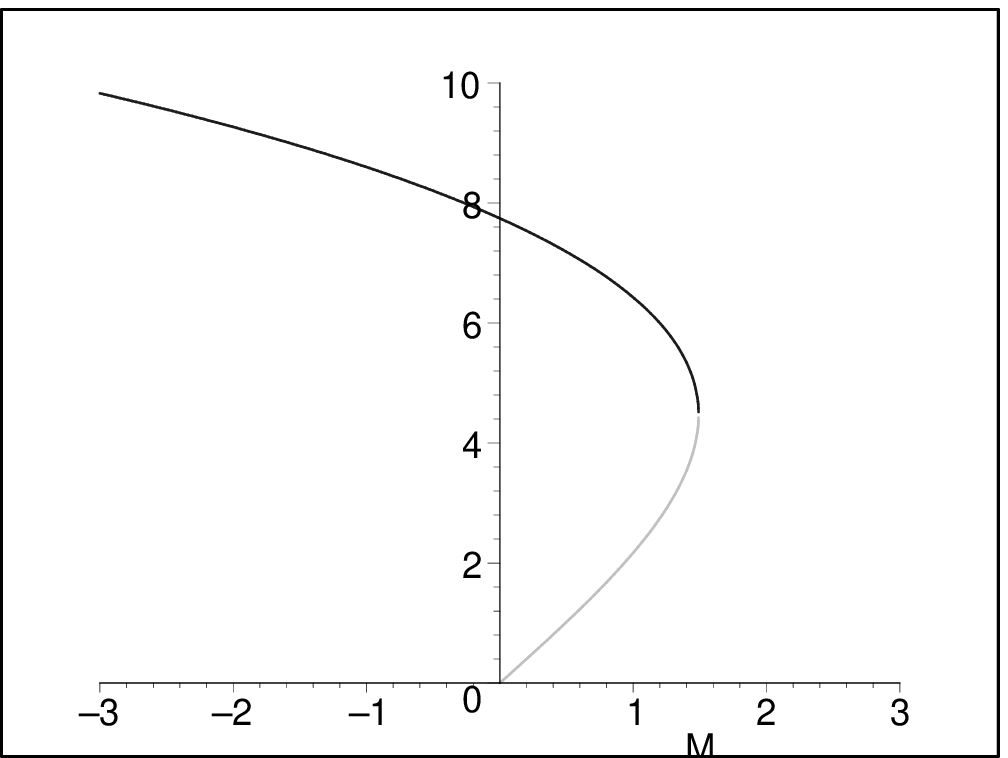}
%{Horizon_k_positive_GR01.eps}
\caption{Plots of $r_+$ (top curve) and $r_-$ (bottom curve) as a function of $M$. These are the plots for $\La=-0.05, k=-1$ (left) and $\La=0.05, k=+1$ (right).}
\label{fig:all2}
\end{figure}

For the case $k=-1$, {\it i.e.,} the hyperbolic membrane case, the horizon structure of the $M>0$ case is similar to the planar case: a black hole horizon exists for $\Lambda<0$ and there is no horizon for $\Lambda\geq 0$.
\footnote{Note that this corresponds to shifting of $f(r)\ra f(r)-1$ in Fig.
\ref{fig:all} (left).}
The situation is quite different for the $M<0$ case (Fig. \ref{fig:all}
(center) and Fig. \ref{fig:all2} (left)), where a membrane solution without
naked singularities is possible for $\Lambda<0$ provided
$|\Lambda|^{-1}<9 M^2$ \ci{Mann:9705}, with inner and outer black membrane
horizons sitting at $r_-$ and $r_+$ respectively (with $r_- < r_+$)
\begin{\eq}
r_+&=&\sqrt{\f{3}{|\Lambda|}} \mbox{cos}\left(\f{1}{3} \mbox{arcsin}  \sqrt{9 M^2 |\Lambda|}\right)-\sqrt{\f{1}{|\Lambda|}} \mbox{sin}\left(\f{1}{3} \mbox{arcsin} \sqrt{9 M^2 |\Lambda|})\right)\, , \no \\
r_-&=&\f{2}{ \sqrt{|\Lambda|}} \mbox{sin}\left(\f{1}{3} \mbox{arcsin} \sqrt{9 M^2 |\Lambda|})\right)\, .
\label{horizon:Mann}
\end{\eq}
On the other hand for the case $\Lambda \geq0$ ($M<0$), there is a
single horizon at $r_-$ but the singularity at $r=0$ is naked
in the causal region $0 \leq r <r_-$, as in the planar case.

Finally, the case $k=+1$, {\it i.e.,} spherical horizon, is the well-known
black hole solution. For $M>0$ (Fig. \ref{fig:all} (right) and
Fig.\ref{fig:all2} (right)), there is a single horizon for $\Lambda\leq0$
located at
\begin{\eq}
r_+&=&
\left\{
\begin{array}{ll}
2 M
& \,\,\mbox{for}~\Lambda=0\, , \\
\left(-3 M +\sqrt{|\Lambda|^{-1}+9 M^2} \right)^{1/3} \Lambda^{-1/3}
+
\left(-3 M +\sqrt{|\Lambda|^{-1}+9 M^2} \right)^{-1/3} \Lambda^{-2/3}
&\,\,\mbox{for}~\Lambda<0\, ,
\end{array}
\right.
\end{\eq}
while for $\Lambda>0$ there are black hole and cosmological horizons at $r_{-}$ and $r_+$ respectively, as given in (\ref{horizon:Mann}) provided $|\Lambda|^{-1}<9 M^2$.

The basic difference between the dS black hole in the last case
($k=+1, \Lambda>0, M>0$) and the case of negative mass, black hyperbolic
membrane ($k=-1, \Lambda<0, M<0$) is that $r_{\mp}$ is its
black hole/cosmological horizon for the former case, while the inner/outer
black membrane horizons of the black hyperbolic membrane, without a
cosmological horizon, for the latter. The case $|\Lambda|^{-1}=9 M^2$ is
the instance that the two horizons coincide and Hawking temperature for the
black hole horizon $r_-$, given by
\begin{\eq}
T_H=\left(\f{\hbar}{4 \pi} \right) \left. \left(\f{df}{dr}\right)\right|_{r_-}
=\left(\f{\hbar}{2 \pi} \right) \left(-\f{ \Lambda r_-}{3}  +\f{ M}{r_-^2}\right)\,,
\end{\eq}
vanishes and matches with that of cosmological horizon $r_+$, such that a thermal equilibrium is reached (Nariai solution) for the former, while (positive) Hawking temperature for the negative mass black hyperbolic membrane vanishes (extremal black brane) for the latter.

Summarizing this section, there are two possible black membrane solutions for
$k=0, M>0$, or $k=-1, M<0$ without naked singularities for $\Lambda<0$.
However, if we consider our current universe as a dS-like space, as implied
by the current accelerating expansion \ci{Ries}, these membrane solutions may not be quite relevant to it. If this is the case, the relevant black membrane solutions may not exist in pure Einstein gravity without matter.

\section{The D=4 Black Hole and Membrane Solutions in IR-modified Ho\v{r}ava Gravity}

\subsection{IR-modified Ho\v{r}ava Gravity and GR Flow Limit Without Fine Tuning}

In order to study Ho\v{r}ava gravity, we write the geometry in terms of
its ADM decomposition
\begin{\eq}
ds^2=-N^2 c^2 dt^2+g_{ij}\left(dx^i+N^i dt\right)\left(dx^j+N^j\,
dt\right)\
\label{ADM}
\end{\eq}
and the IR-modified Ho\v{r}ava action then reads
\begin{\eq}
S &= & \int dt d^3 x
\sqrt{g}N\left[\frac{2}{\kappa^2}\left(K_{ij}K^{ij}-\lambda
K^2\right)-\frac{\kappa^2}{2\nu^4}C_{ij}C^{ij}+\frac{\kappa^2
\mu}{2\nu^2}\epsilon^{ijk} R^{(3)}_{i\ell} \nabla_{j}R^{(3)\ell}{}_k
\right.
\nonumber \\
&&\left. -\frac{\kappa^2\mu^2}{8} R^{(3)}_{ij}
R^{(3)ij}+\frac{\kappa^2 \mu^2}{8(3\lambda-1)}
\left(\frac{4\lambda-1}{4}(R^{(3)})^2-\Lambda_W R^{(3)}+3
\Lambda_W^2\right)+\frac{\kappa^2 \mu^2 \om}{8(3\lambda-1)}
R^{(3)}\right]\ , \label{horava}
\end{\eq}
where
\begin{\eq}
 K_{ij}=\frac{1}{2N}\left(\dot{g}_{ij}-\nabla_i
N_j-\nabla_jN_i\right)\,
 \end{\eq}
is the extrinsic curvature,
\begin{\eq}
 C^{ij}=\epsilon^{ik\ell}\nabla_k
\left(R^{(3)j}{}_\ell-\frac{1}{4}R^{(3)} \delta^j_\ell\right)\,
 \end{\eq}
is the Cotton tensor, and $\kappa,\lambda,\nu,\mu, \La_W$, and $\om$ are
coupling constants. From the higher spatial derivatives up to six orders,
the theory becomes power-counting renormalizable with the dimensionless couplings $\kappa$ and $\nu$. The last term in the action represents a ``soft" violation of the detailed balance condition \ci{Hora,Keha,Nast,Park:0905} that modifies the IR behavior without changing the improved UV behavior. Notice that, being the action non-symmetric in space and time, it is crucial that the metric (\ref{ADM}) has the right signature, with time-like coordinate $t$ and space-like coordinate $x^i$, so that the original Ho\v{r}ava reasoning on renormalizability is valid. This determines $t$ as the time coordinate uniquely, in contrast to GR case.

Naively, one might expect that Ho\v{r}ava gravity would reduce to GR
by assuming higher-derivative terms are negligible at large distances,
{\it i.e.,} low energy, but there are some subtleties involved. For example,
the truncated theory, which is effective at large distances, has a different
constraint structure than that of the full theory \cite{Hora:08,Li}. So in
order to recover GR, we consider the more general limiting procedure which
entails the flow of the coupling constants as well as that of the
characteristic length scale. Actually, we find that in order to recover GR,
the coupling constants need to flow as
\begin{\eq}
\lambda \ra 1, ~~
\qquad
\mu \ra 0, ~~
\qquad
\nu \ra \infty,~~
\qquad
\Lambda_W \ra \infty\, ,
\label{GR_limit_1}
\end{\eq}
with
\begin{\eq}
\mu^2 \Lambda_W^2 \sim \mbox{fixed},~~
\quad\qquad\qquad
\mu^2 \omega\sim \mbox{fixed}.
\label{GR_limit_2}
\end{\eq}
In this flow, all the higher spatial derivative terms and the term
proportional to $\mu^2 \Lambda_W R^{(3)}$ vanish, and only the kinetic,
cosmological constant, and IR-modification terms remain. We note that this
kind of consistent flow is not possible in the original Ho\v{r}ava model
with $\om=0$, without introducing a hypothetical fine-tuning mechanism
in order to subtract an infinite constant and get a finite cosmological
term \ci{Lu}. Now, by comparing with the Einstein-Hilbert action
(recovering the speed of light $c$) \ci{Park:0906,Ryde},
\begin{\eq}
S_{\rm EH} =  \f{c^4}{16 \pi G} \int dt d^3 x
\sqrt{g}N\left[\frac{1}{c^2}\left(K_{ij}K^{ij}- K^2 \right) +R^{(3)}-\f{2 \La}{c^2} \right],
\nonumber
\end{\eq}
one can obtain the following relations for the fundamental parameters of GR \footnote{These relations generalize those of \ci{Keha} for $\omega=8
\mu^2(3\lambda-1)/\kappa^2$ to an arbitrary $\om$ and non-vanishing $\La_W$,
but they differ from the original ones \ci{Hora,Lu,Park:0905,Park:0906}.},
\begin{\eq}
c^2=\f{\kappa^4 \mu^2 \omega}{32},
\qquad \quad
G=\f{\kappa^2 c^2}{32 \pi},
\qquad \quad
\Lambda=-\f{3\Lambda_W^2 c^2}{2\omega}\, .
\label{GR_limit_3}
\end{\eq}
These relations imply that $\mu^2 \omega >0$ and $\kappa^2>0$ from the physical conditions of $c^2>0$ and $G>0$. The AdS and dS space in Einstein gravity limit can be described, with $\Lambda_W^2>0$, by $\omega>0, \mu^2>0$ and
$\omega<0, \mu^2<0$ respectively, which are degenerate in the flat space case
$\Lambda_W=0$. Notice that these relations cannot be defined in the original Ho\v{r}ava model with $\om=0$ \ci{Hora}; this means that the $\om=0$ case does not have a straightforward way to compare with our universe.

\subsection{Metric Ansatz and General Solution with IR Lorentz Invariance ($\la=1$)}
Let us consider now the static and maximally symmetric solution with the
metric ansatz (\ref{ansatz})-(\ref{sub_metric}). By substituting it into the
action (\ref{horava}), the resulting reduced Lagrangian, after angular
integration, is given by
\begin{\eq}
{\cal{L}}&=&\frac{\kappa^2\mu^2 \Om_k}{8(1-3\lambda)}\frac{N}{\sqrt{f}}
\left[(2\lambda-1)\frac{(f-k)^2}{r^2}
-2\lambda\frac{f-k}{r}f'+\frac{\lambda-1}{2}f'^2
-2 (\om-\La_W) (k-f-rf') - 3 \La_W^2 r^2 \right]\ .
\end{\eq}
The resulting equations of motion are
\begin{\eq}
&&(2\lambda-1)\frac{(f-k)^2}{r^2}-
2\lambda\frac{f-k}{r}f'+\frac{\lambda-1}{2}f'^2-2 (\om-\La_W)
(k-f-rf')- 3 \La_W^2 r^2 =0\, ,
\nonumber \\
&& \left(\frac{N}{\sqrt{f}}\right)' \left((\lambda-1)f'-2\lambda
\frac{f-k}{r}+2(\om-\La_W)
r\right)+(\lambda-1)\frac{N}{\sqrt{f}}\left(f''-\frac{2(f-k)}{r^2}\right)=0\, ,
\end{\eq}
obtained by varying the functions $N$ and $f$, respectively.

For the $\lambda=1$ case, which reduces to the standard Einstein-Hilbert action in the IR limit (so that there is no Lorentz violation in IR), the solutions
are obtained as \ci{Park:0905}
\begin{\eq}
N^2=f=
k+(\om-\La_W) r^2 +\ep \sqrt{r[\omega (\om-2 \La_W) r^3 + \be]}\, ,
\label{solution}
\end{\eq}
where $\ep=\pm 1$ and $\be$ is an integration constant \footnote{If one adds another IR-modification term ${\kappa^2 \mu^2}({8 (3 \la-1)})^{-1} \hat{\be} \La_W^2$ as in \ci{Hora,Nast},  the solution becomes $N^2=f=k+(\om-\La_W) r^2 + \ep \sqrt{r[\{\omega (\om-2 \La_W)+ \hat{\be} \La_W^2/3 \} r^3 + \be]}$\,. This can be obtained by redefining the parameters $\La_W \ra \sqrt{1-\hat{\be}/3}~ \La_W,~ \om \ra \om + (\sqrt{1-\hat{\be}/3}-1) \La_W$ in (\ref{solution}). This is also true at the action level so that the IR-modification term in (\ref{horava}) is more or less unique.}.

In the ``asymptotic" region $r \gg [\be /\om ( \om-2 \La_W)]^{1/3}$, the
above solution behaves as
\begin{\eq}
N^2=f=k+\left(\om-\La_W + \ep |\om| \sqrt{1-\f{2\La_W}{\om} } \right) r^2 + \f{\ep \beta}{2 |\om| \sqrt{1-2 \La_W/\om}}\f{1}{ r} + {\cal O}(r^{-4})\,,
\label{approx}
\end{\eq}
but, as we see, this is not enough to get the conventional results of GR.
Now then, by defining a new parameter $M$ as $\be=4 \om M$ and considering the
``GR" limit
\footnote{From (\ref{GR_limit_2}), one obtains
$|\om | \sim |\La_W^2| \gg |\La_W|$ as $|\La_W| \ra \infty$.}
$|\om| \gg |\La_W|$, this becomes
\begin{\eq}
N^2=f=k-\f{\La }{3 c^2 } r^2 - \f{2M}{r}
+ {\cal O}(r^{-4})\, ,
\label{f_GR_limit}
\end{\eq}
in agreement \footnote{There were some, unexplained, factor disagreements in the GR limit of (\ref{approx}) with Schwarzschild-AdS/dS black hole for the original definition \ci{Hora,Lu,Park:0905}. Now with the new definitions of
(\ref{GR_limit_3}), this problem does not occur and we have perfect agreement up to order $r^{-4}$. } with the standard Schwarzschild-AdS/dS black hole
(\ref{GR_solution}), by taking $\ep=-1$ for the AdS/flat case ($\La \leq0$ or equivalently $\om>0$), and $\ep=+1$ for the dS case ($\La>0$ or equivalently $\om<0$). Since we are interested in solutions to IR-modified Ho\v{r}ava gravity that flow into GR solutions under the IR limit, hereafter we only consider the $\ep=-1~ (+1)$ branch for $\om>0~ (<0)$ which we call the ``AdS(dS) branch". This shows the importance of the GR flow, which achieves a peculiar form of flows of coupling constants as in (\ref{GR_limit_1}-\ref{GR_limit_3}), as well as the asymptotic limit in order to recover the results of GR. In other words:

~
\begin{center}{\it ``The GR limit of the Ho\v{r}ava black hole/membrane is not reached in the asymptotic region generically, but only in
the $\Lambda_W=0$ case".}
\end{center}
~

\noindent This may explain the significant difference between the Schwarzschild-AdS/dS solution and the L\"u, Mei, Pope's solution \ci{Lu} of the original Ho\v{r}ava gravity with the detailed balance condition. For the latter, the GR limit cannot be defined in the asymptotic region due to the absence of $\om$, implying that there is no way to compare to our universe.

\subsection{Unusual Singularities and Horizons}
The solution (\ref{solution}) has a spatial curvature invariant
\begin{\eq}
R^{(3)}&=&-6 \left( (\om-\La_W) +\ep \f{\om (\om-2 \La_W) r^3 + \be/2}{r \sqrt{r[\omega (\om-2 \La_W) r^3 + \be]}}\right)\, .
\label{curvature}
\end{\eq}
This shows that, in the asymptotic limit, the solution behaves as a constant
curvature space $R^{(3)}\approx -6 [(\om-\La_W) +\ep \sqrt{\om(\om-2 \La_W)}]$
flowing into an asymptotically AdS/dS space-time $R^{(3)}\approx 2 \La$ in the GR limit (\ref{GR_limit_3}).

For $\beta>0$, the usual point-like curvature singularity at $r=0$ is
present as in Einstein gravity, but now with a  milder form
$R^{(3)}_{ij}R^{(3)ij}\approx (27/8)\be r^{-3},~
R^{(3)}_{ijkl} R^{(3) ijkl}=4 R^{(3)}_{ij}R^{(3)ij}-{R^{(3)}}^2
\approx 9  \be  r^{-3}$ in contrast to the $r^{-6}$ of the GR case
(\ref{singularity:GR}). On the other hand, when $\om (2 \La_W-\om) \neq 0$
and $\be \neq 0$, the above expression shows an unusual surface-like
curvature singularity \footnote{A similar surface singularity at
$r_{\La}\equiv (-2 M/\La_W)^{1/3}$ has been found for the projectable form of
the AdS-\Sch black hole solution $(\La_W<0)$ \ci{Lu},
$ds^2=-dt^2+(dr +\sqrt{M/r+\La_W r^2/2}~ dt)^2 + r^2 d \Om$ with the detailed
balance ({\it i.e}, $\om=0$ in our context) \ci{Cai:1001}. In contrast to our
non-projectable solutions, the projectable solution does not have
contributions from higher-spacial derivatives due to ``flat"-spatial metric
$g_{ij}$, {\it i.e.,} $R^{(3)}=0$, and the singularities are captured by
the extrinsic curvature scalar $K$, instead. In order to compare $r_{\La}$
with $r_S$ in our case, one might try to consider $\om=0$ with $\be=4 \om M$
and obtains $r_S=(2 M/\La_W)^{1/3}$, which {\it disagrees} with $r_{\La}$.
But in this case, $R^{(3)}$ in (\ref{curvature}) is subtle in identifying
the singularity at $r_S$ so that the direct connection between the
singularities for the two distinct solutions is not quite manifest.},
sitting at
\begin{\eq}
r_S=\left( \f{\be}{\om (2 \La_W-\om)}\right)^{1/3}\, , \label{ring_singularity}
\end{\eq}
where the denominator of the second term in (\ref{curvature}) vanishes and
$R^{(3)} \approx \sqrt{3} \ep \be^{1/3} [\omega (2 \La_W-\om) ]^{1/6}
(r-r_S)^{-1/2}$ near $r= r_S$, with a lower degree of divergence than the
aforementioned point-like singularity. For the case $\om (2 \La_W-\om) = 0$,
only the point singularity at $r=0$ survives, with no additional singularity in the curvature invariants. Note that these singularities are physical ones which cannot be removed by coordinate transformations in the group of foliation preserving diffeomorphism.

In what follows, due to the singularity at $r=0$, we consider the range $r>0$
without loss of generality, as in the previous section for the GR case.
Moreover, since the surface-like singularity sits exactly at the
location where the square-root term in (\ref{solution}) vanishes,
the metric becomes complex-valued beyond $r_S$. This implies that we
must consider the ranges
\begin{\eq}
\left\{
\begin{array}{ll}
r>r_S >0&\qquad\mbox{for}~ \be<0,~ \om (2 \La_W-\om)<0\, ,
\\ &~ \\
0<r<r_S& \qquad\mbox{for}~ \be>0,~ \om (2 \La_W-\om)>0\,
\end{array}
\right.
\label{lachota}
\end{\eq}
in order to have a real-valued metric.

From the metric ansatz (\ref{ansatz}), we define the ``observer region'' of
our solution as that where $t$ is the time and $r$ is the space, in other
words, $N^2(r),f(r)>0$ so that measurements can be made by a fixed observer.
Note that only in this region the power-counting renormalizable Ho\v{r}ava
theory is correctly defined \footnote{In the regions of our solutions where
$N^2(r),f(r)<0$, the signature of the metric is such that $r$
becomes the time variable. Since our system has higher $r$ derivatives,
Ostrogradsky ghost might appear there, and the solution become unstable. A
definitive answer would require a deeper investigation that we plan to pursue
somewhere else \cite{inProgress}. For the time being, the above defined
observer regions of our solutions can be regarded as building blocks to
construct wormhole-like metrics, like those of
\cite{Bott:0906, Bellorin:2014qca, Bellorin:2015oja}, in which
$N^2(r),f(r)>0$ everywhere.}. So, from the point of the observer region, the singularities should be avoided or hidden behind horizons \footnote{The notion of the horizon is an emergent concept at low energy and so the cosmic censorship may be violated at higher energies. But here, we adopt the cosmic censorship as an emergent notion at low energy also. We will discuss more about this point in Sec. VI. and in a forthcoming publication \cite{inProgress}}.

~

Now, in order to find the horizons, we note that the horizon condition $N^2=f=0$ can be rewritten as $k+(\om-\La_W) r^2 =-\ep \sqrt{r[\omega (\om-2 \La_W) r^3 + \be]}$. By squaring both sides, we need to solve the polynomial $\Lambda_W^2r^4+2k(\omega\!-\!\Lambda_W)r^2-\beta r+ k^2$ to obtain its positive zeros, and then filtering them with the additional requirement that the sign of $k+(\omega-\Lambda_W)r^2$ at the zero has to be $-\epsilon$.
In this way, one can obtain the two roots, generically
\begin{\eq}
r_\pm &=&S \pm\f{1}{|\Lambda_W|}\sqrt{\f{\beta}{4S}-S^2\Lambda_W^2-{k (\omega-\Lambda_W)}}\, ,
\label{root}
\end{\eq}
where
\begin{\eq}
&&S=\sqrt{-\f{k(\omega-\Lambda_W)}{3\Lambda_W^2} +\f{1}{12\Lambda_W^2} \left(Q+\f{\Delta_0}{Q} \right)},\qquad
\qquad
\ \
Q=\left( \f{\Delta_1 +\sqrt{ \Delta_1^2 -4 \De_0^3}}{2}\right)^{\frac13}\, ,   \\
&&
\no\\&&
\De_0 %=c^2+12ae
=4 k^2 (\omega^2-2 \omega \Lambda_W +4 \Lambda_W^2), \qquad\qquad \ \ \ \,
\De_1 %=2 c^3+27a d^2 -72 a c e
=16 k^3 (\omega-\Lambda_W)^3+27 \Lambda_W^2 \beta^2-144 \Lambda_W^2 k (\omega-\Lambda_W)\, .    \no
\end{\eq}
Notice that the roots $r_\pm$ above are not necessarily positive nor real, so they may not represent real horizons for some range of parameters. We will explore this issue in the forthcoming sections. The number of horizons at positive $r$ can also be obtained by making use of Descartes' rule of signs in the aforementioned polynomial.

The roots $r_\pm$ may coincide at $r=r_*$ when there is a double root,
{\it i.e.,} $f(r_*)=f'( r_*)=0$, which can be solved for $\beta$ obtaining
\begin{\eq}
\beta_{*}=\frac{4 }{3\sqrt{3} |\Lambda_W |}
\sqrt{
-k(\omega-\La_W)+|k|\sqrt{(\omega ^2-2 \omega \Lambda_W +4 \Lambda_W^2)}}
\left( 2 k(\omega-\La_W)+|k|\sqrt{(\omega ^2-2 \omega \Lambda_W +4 \Lambda_W^2)}\right),
\label{beta_extrem}
\end{\eq}
as the minimum value of the integration constant $\beta$, for a given $\Lambda_W$. The reason of a minimum value $\beta_*$ can also be understood
from the black hole/membrane thermodynamics (see Sec. IV). We can also solve
for $r_*$ to get
\begin{\eq}
r_* &=&
\sqrt{\frac{-k(\omega
   -\Lambda_W)+|k|\sqrt{\omega ^2-2 \omega \Lambda_W +4 \Lambda_W^2}}{3{\Lambda_W^2}}
   }\, .
\label{dop}
\end{\eq}

On the other hand, the horizons at $r_{+}$ or $r_-$ may coincide with
the surface-like singularity at ${\tilde r}_{\pm}$ when
$\beta=\tilde{\beta}$ with
\begin{\eq}
\tilde{r}_{\pm}=\sqrt{\f{k}{\La_W-\om}},
\qquad
\qquad
\tilde{\be}=\om (2 \La_W- \om) \left( \f{k}{\La_W-\om}\right)^{3/2},
\end{\eq}
which can be obtained, by solving $r_S=r_{\pm}\equiv \tilde{r}_{\pm}.$

\subsubsection{Flat membrane solution $(k=0)$}

We first consider the flat membrane case $k=0$ and classify the solutions according to the sign of $\beta$.

~

\noindent{\it A. Case $\beta>0$}:

~

In this case, there is an inner horizon, for any {\it finite} $\om$, at
\begin{\eq}
r_-=0\, ,
\end{\eq}
so that the point singularity at $r=0$ is not naked unless we consider the
trivial case of Minkowski vacuum, $\be=\La_W=0$. This is a genuine effect of
the higher-derivative terms in Ho\v{r}ava gravity which is absent in the GR.
Moreover, there is an outer horizon at
\begin{\eq}
r_+=\left( \f{\be}{\La_W^2}\right)^{1/3}\, ,
\end{\eq}
which exists in the AdS branch ($\ep=-1$) of the solution (\ref{solution})
for $\om>\La_W$ and in the dS branch ($\ep=+1$) for $\om <\La_W$.
Interestingly, this outer horizon is exactly the same as
(\ref{horizon_GR_flat}) in Einstein gravity, with the identification of
$\be=4 \om M$ and $\La$ as in (\ref{GR_limit_3}) and the solution is similarly interpreted as a black plane.

\begin{figure}
\includegraphics[width=7.3cm,keepaspectratio]{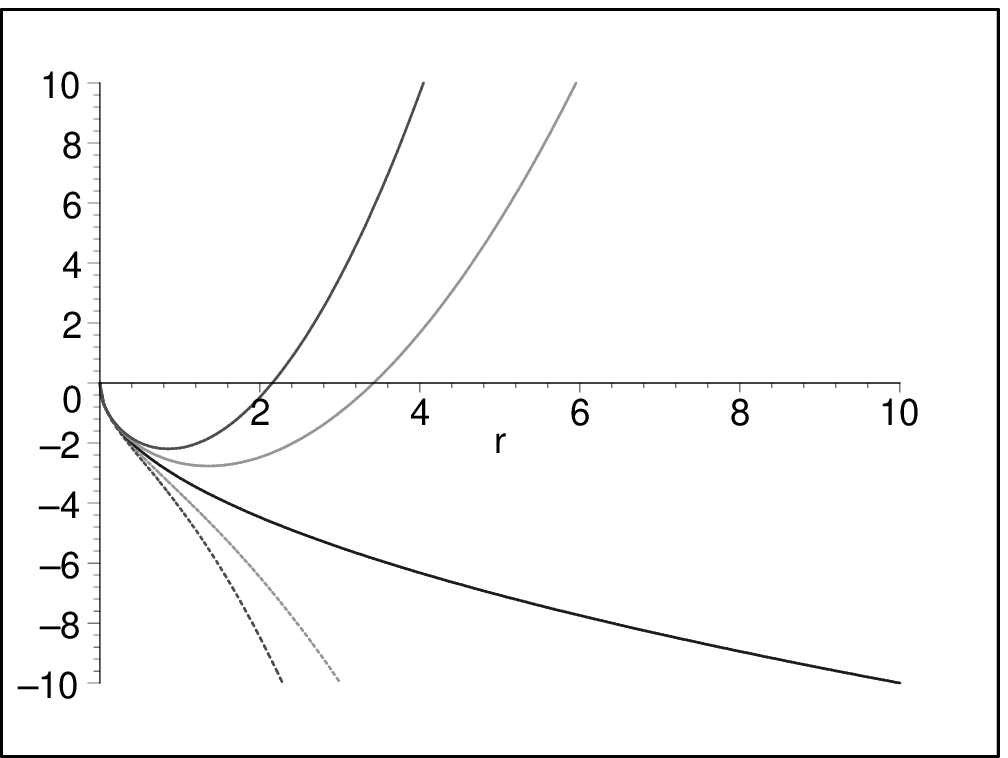}
%{k_zero_omega_zero_AdS01.eps}
\qquad \qquad
\includegraphics[width=7.3cm,keepaspectratio]{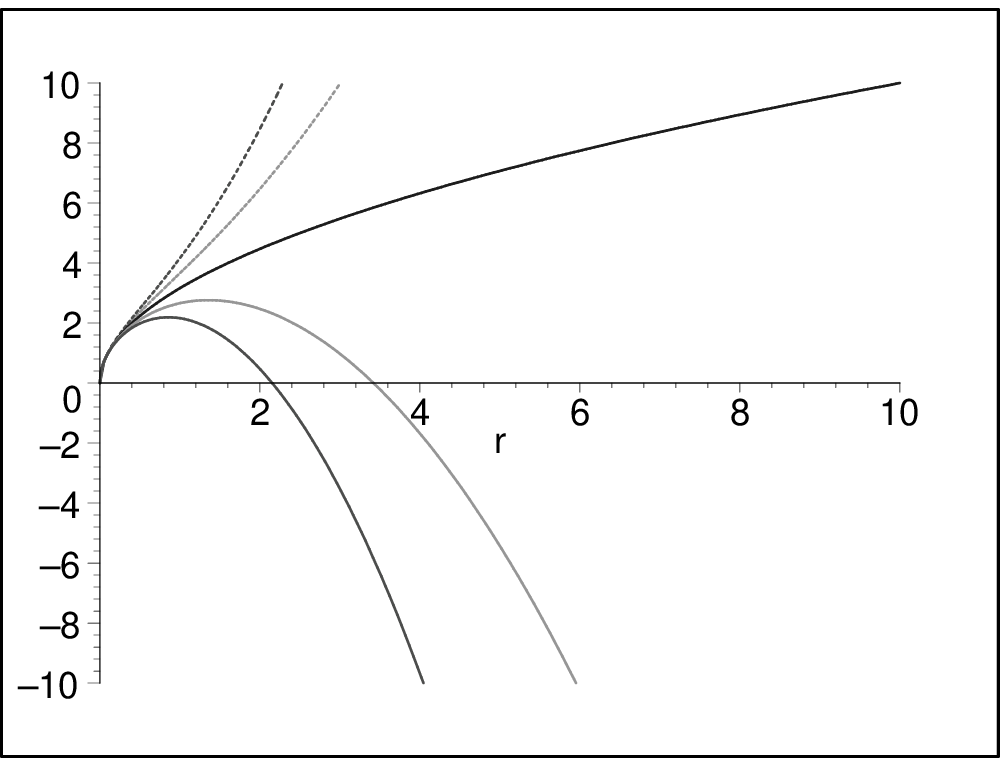}
%{k_zero_omega_zero_dS01.eps}
\caption{Plots of $f(r)$ for $k=0$, $\beta>0$, $\omega=0$. \underline{Left}: Case $\ep=-1$. We have plotted for $\La_W=-1, -0.5, 0, 0.5, 1~(\beta=10) $ from top to bottom, respectively. The cases $\Lambda_W<0$ correspond to the black planes in AdS space with the horizon at $r_-=0$ and $r_+=(\be/\La_W^2)^{(1/3)}$. The cases $\La_W \geq 0$ have a horizon at $r_-=0$ so that the singularity is not naked, but they have no region in which the variable $r$ is space-like. \underline{Right}: Case $\ep=+1$. We have plotted for $\La_W=1, 0.5, 0, -0.5, -1~(\beta=10)$ from bottom to top, respectively.  The cases $\Lambda_W>0$ show the cosmological horizon in dS space at $r_+=(\be/\La_W^2)^{(1/3)}$ as well as the black plane horizon at $r_-=0$ so that the $r=0$ singularity is not naked. The cases  $\La_W \leq 0$ have a horizon at $r_-=0$ so that singularity is not naked, and the variable $r$ has the space-like signature in its whole range of $r>0$.}
\label{fig:Horava_horizon_k_zero_omega_zero_AdSanddS}
\end{figure}

\begin{figure}
\includegraphics[width=7.3cm,keepaspectratio]{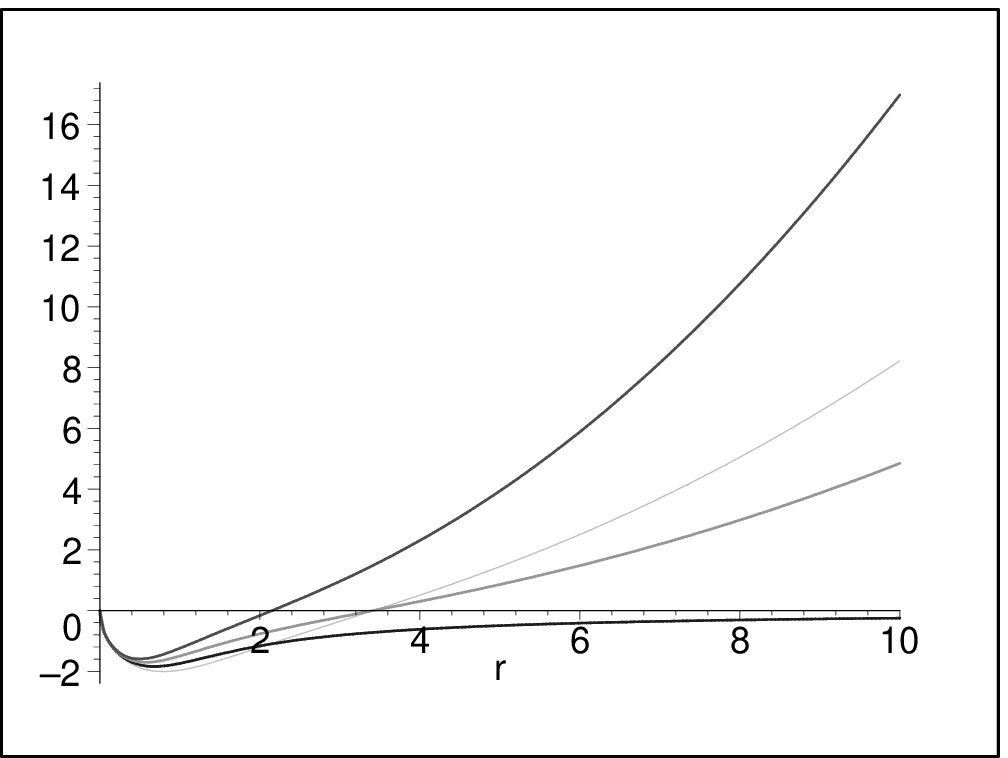}
%{k_zero_omega_positive01.eps}
\qquad \qquad
\includegraphics[width=7.3cm,keepaspectratio]{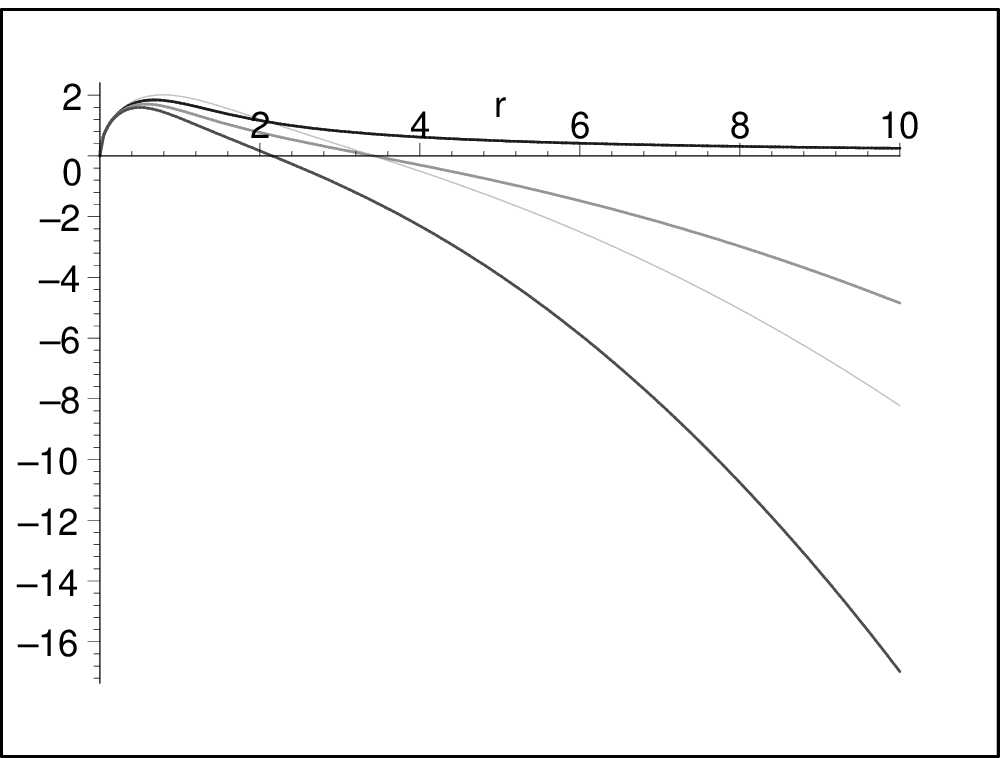}
%{k_zero_omega_negative_plus01.eps}
\caption{Plots of $f(r)$ for $k=0$, $\beta>0$, $\om (2 \La_W-\om)<0$.
\underline{Left}: Case $\ep=-1$ $\om >0$, $\omega>2\La_W$. The three (thick) curves denote black membranes for $\La_W=-1,-0.5,0, \om=2~(\beta=10)$ from top to bottom, respectively. The behaviors are qualitatively the same as the case (a) with $\La_W<0$. This property extends also to $\La_W>0$ region, if $\om > 2 \La_W$ is satisfied, as plotted by thin curve for $\La_W=0.5$. \underline{Right}: Case $\ep=+1$, $\om<0$, $\omega<2 \La_W$. The two (thick) curves denote black membranes for $\La_W=1, 0.5, \om=-2 $ from top to bottom, respectively. The behaviors are qualitatively the same as the case (b) with $\La_W>0$. This property extends also to $\La_W<0$ region, if $\om < 2 \La_W$ is satisfied, as plotted by thin curve for $\La_W=-0.5$.} \label{fig:Horava_horizon_k_zero_omega_positive_plus_and_omega_negative_plus}
\end{figure}

\begin{figure}
\includegraphics[width=7.3cm,keepaspectratio]{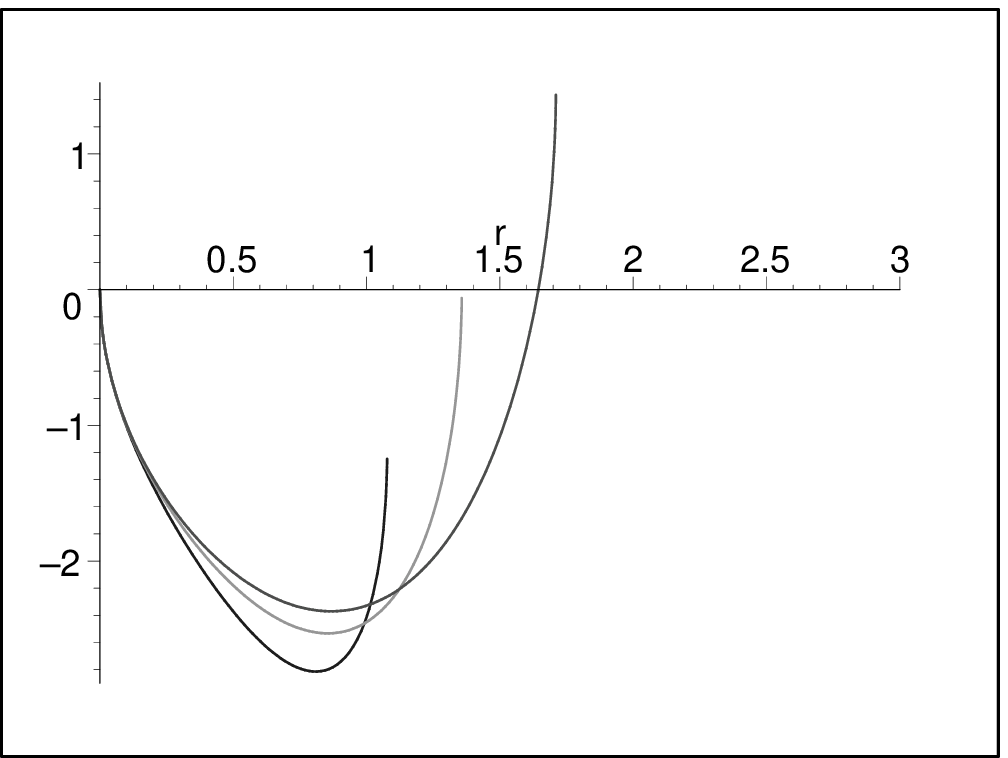}
%{k_zero_omega_positive_complex01.eps}
\qquad \qquad
\includegraphics[width=7.3cm,keepaspectratio]{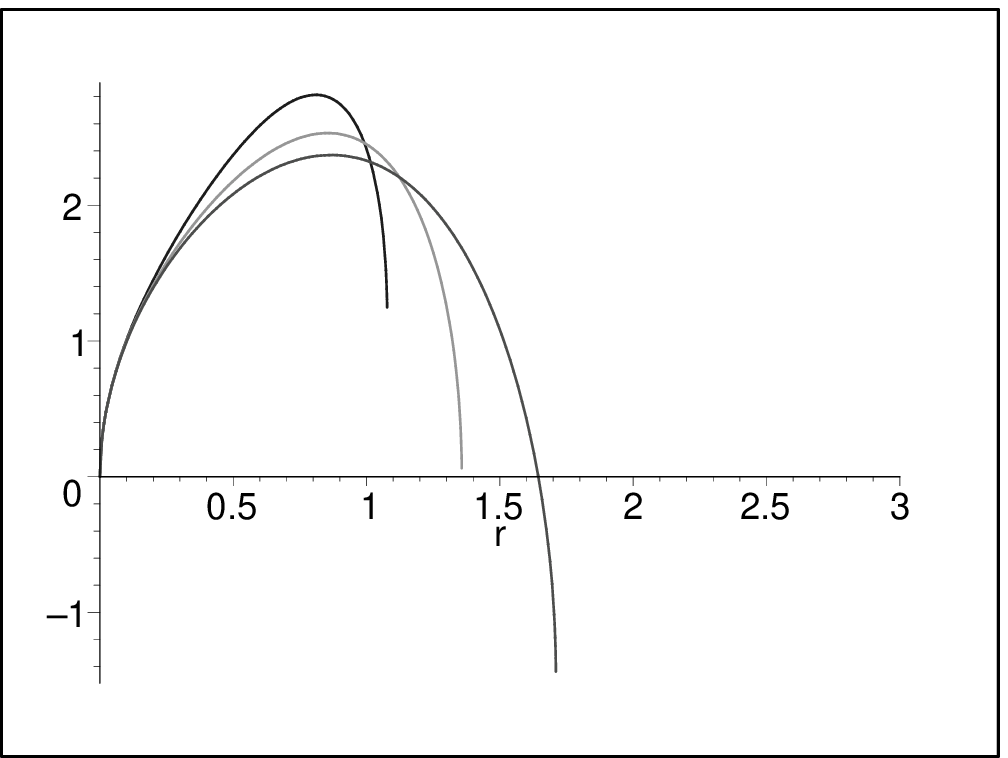}
%{k_zero_omega_negative_plus_except01.eps}
\caption{Plots of $f(r)$ for $k=0$, $\beta>0$, $\om (2 \La_W-\om)>0$. \underline{Left}: Case $\ep=-1$, $0< \om < 2 \La_W$. The three curves denote black membranes for $\La_W=3,2,1.5,\om=2 ~(\beta=10)$ from bottom to top, respectively. For $\om < \La_W$, there is a curvature singularity at $r_S$ where the curve ends and beyond which it becomes complex valued, but {\it no} black hole horizon $r_+$ (bottom curve). On the other hand, for $\om=\La_W$ (middle curve) and $\om > \La_W$ (top curve), $r_S$ is located {\it on} or {\it outside} the horizon $r_+$. \underline{Right}: Case $\ep=+1$, $2 \La_W < \om  <0$. The curves denote black membranes for $\La_W=-3,-2,-1.5, \om=-2 ~(\beta=10)$ from top to bottom, respectively. For $\om > \La_W$, there is a curvature singularity at $r_S$ where the curve ends and beyond which it becomes complex valued, but {\it no} cosmological horizon $r_+$ (top curve). On the other hand, for $\om=\La_W$ (middle curve) and $\om< \La_W$ (bottom curve), $r_S$ is located {\it on} or {\it behind} the cosmological horizon $r_+$.} \label{fig:Horava_horizon_k_zero_omega_positive_complex_and_omega_negative_plus_except}
\end{figure}

Now, in order to see whether the surface-like curvature singularity
(\ref{ring_singularity}) is naked in our observer region or not, one might try to consider the condition for $r_+ \ge r_S$ if $r_+$ and $r_S$ exist. Supposing that $r_+ \ge r_S$ implies $\La_W^2-\om (2 \La_W -\om) =(\La_W - \om)^2 \le 0$, but this is impossible if $\Lambda_W\neq\omega$, and the case $\La_W=\om$ is when $r_+=r_S$. This proves that $r_+ \le r_S$ if $r_S$ exists. According to (\ref{lachota}), the surface-like singularity exists for $\beta>0$ when
$\omega(2\Lambda_W-\omega)>0$, and in such case the allowed region for the
radial coordinate is $0<r<r_S$. So we can have viable black membrane solutions
without naked singularities either when $r_S$ does not exist, {\it i.e.},
$\omega(2\Lambda_W-\omega)\leq 0$ or when $r_S$ is hidden behind the
cosmological horizon $r_+\le r_S$. Then the possible solutions are (a) $\om=0$, (b) $\om=2 \La_W$, (c) $\om > 0, \om > 2 \La_W$, (d) $\om < 0, \om < 2 \La_W$,
and (e) $2 \La_W < \om \le \La_W <0$. Some more details are in order.\\
\begin{enumerate}[(a)]
\item $\om=0$ : This case corresponds to the plane solution in the original Ho\v{r}ava theory, where the GR is not recovered under the IR flows
(\ref{GR_limit_1})-(\ref{GR_limit_3}) \ci{Argu:1008,Cai}. Here, the
surface-like singularity at $r=r_S$ does not exist, though the horizon at
$r_{-}$ does. The horizon at $r_+$ exists for $\Lambda_W>0$ in the dS
branch ($\epsilon=+1$) as a cosmological horizon, with the observer region
$N^2, f>0$ for $0<r<r_+$, and for $\Lambda_W<0$ in the AdS branch
($\epsilon=-1$) as a black plane horizon, with the observer region for $r>r_+$ (Fig. \ref{fig:Horava_horizon_k_zero_omega_zero_AdSanddS}). However, for the case $\La_W=0$, there is no a priori reason to choose
which of the given two branches, due to the lack of a GR limit.

\item $\om=2 \La_W$ : This case corresponds to $-\La_W  \ra \La_W$ in the result of (a) and so all the properties can be understood just by flipping the sign of $\La_W$ in Fig. \ref{fig:Horava_horizon_k_zero_omega_zero_AdSanddS}.

\item $\om > 0, \om > 2 \La_W, \epsilon=-1$ : In contrast to the cases of (a) and (b), this case reduces to GR under the IR flows (\ref{GR_limit_1})-(\ref{GR_limit_3}). But, similar to the cases of (a) and (b), the surface-like curvature singularity at $r_S$ does not exist, and the $r=0$ singularity is hidden by the coincident inner horizon at $r_-=0$ as well as by the outer horizon at $r_+$, with the
    observer region for $r>r_+$ (Fig. \ref{fig:Horava_horizon_k_zero_omega_positive_plus_and_omega_negative_plus} (left)).
According to the IR limit (\ref{GR_limit_3}) with the identification of
$\be=4 \om M$ as in (\ref{f_GR_limit}), this case flows to the case
$M>0, \La<0$ of GR. However, for the case $\La_W=0$, there is no observer region with the space-like coordinate $r$.

\item $\om < 0, \om < 2 \La_W, \epsilon=+1$: As in the case (c), this case reduces to GR under the IR flows (\ref{GR_limit_1})-(\ref{GR_limit_3}). Again, this case does not confront the surface-like curvature singularity at $r_S$. The horizons at $r_\pm$ exist with the observer region $0<r<r_+$ and the horizon at $r_+$ being a cosmological one (Fig. \ref{fig:Horava_horizon_k_zero_omega_positive_plus_and_omega_negative_plus} (right)). According to the IR limit (\ref{GR_limit_3}) with $\be=4 \om M$ as in (\ref{f_GR_limit}), this case flows into the case $M<0, \La>0$ of GR. Note that the existence of this viable solution is basically due to the existence of an inner horizon $r_-=0$ as a higher-spatial derivative effect that is absent in GR.\\

Except in the above four cases, one can confront the curvature singularity at $r_S$ (Fig. \ref{fig:Horava_horizon_k_zero_omega_positive_complex_and_omega_negative_plus_except}), but there is one interesting viable case.

\item $2 \La_W < \om \le \La_W <0, \ep=+1$: Here, there is a cosmological
horizon at $r_+$ and so the curvature singularity at $r_S$ is always beyond the horizon $r_+ \le r_S$, as has been proven above, and hidden from
the observer region $0<r<r_+$ (Fig.\ref{fig:Horava_horizon_k_zero_omega_positive_complex_and_omega_negative_plus_except}
 (right); middle and bottom curves).
\end{enumerate}
~

\noindent{\it B. Case $\beta<0$}:

~

\begin{figure}
\includegraphics[width=7.3cm,keepaspectratio]{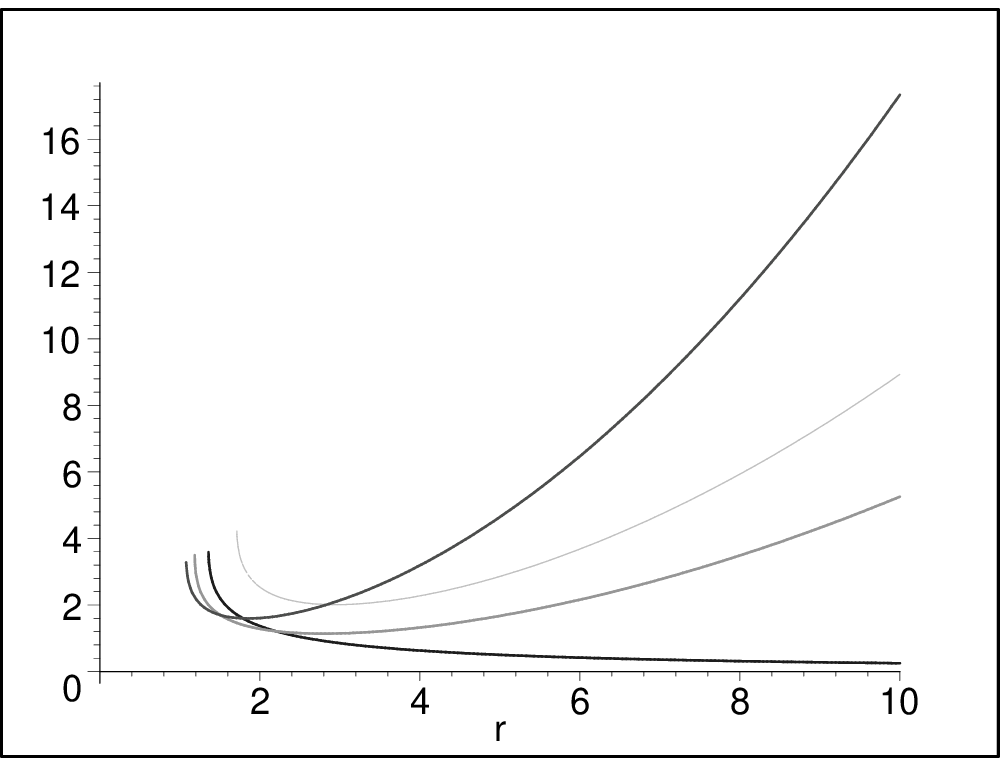}
%{k_zero_omega_positive_complex_left01.eps}
\qquad \qquad
\includegraphics[width=7.3cm,keepaspectratio]{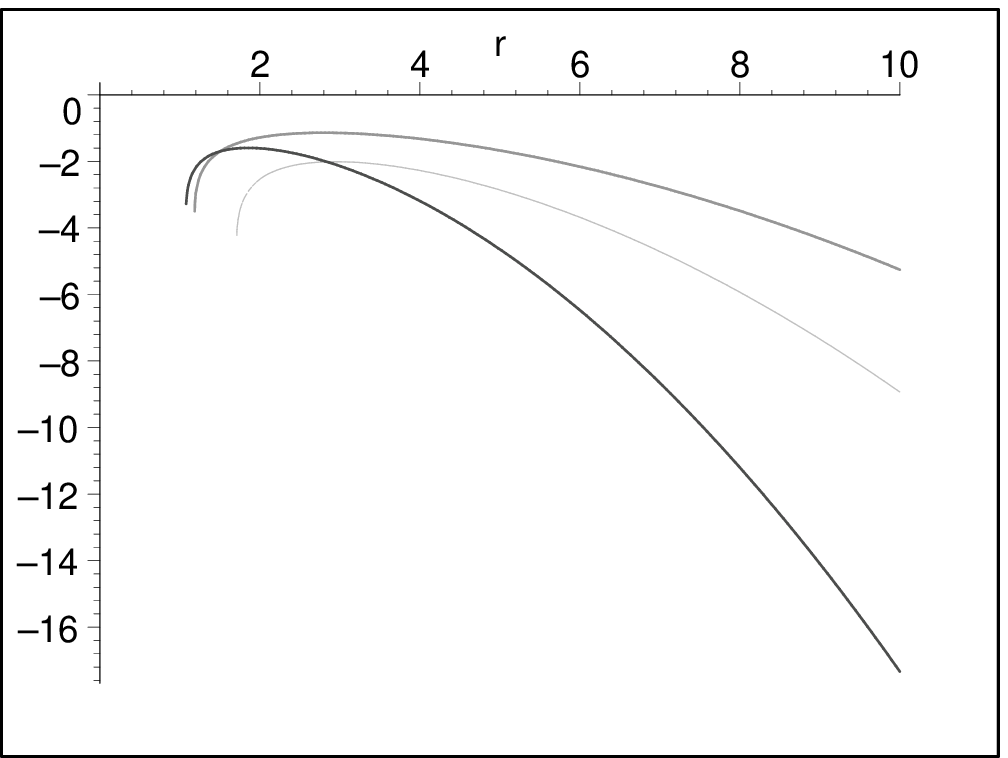}
%{k_zero_omega_negative_plus_complex_left01.eps}
\caption{Plots of $f(r)$ for $k=0$, $\beta<0$, $\om (2 \La_W-\om)<0$.
\underline{Left}: Case $\ep=-1$, $\om > 0$, $\om > 2 \La_W$. The curves denote black membranes for $\La_W=-1,-0.5,0,0.5, \om=2 ~(\beta=-10)$ from left to right, respectively. There is a curvature singularity at $r_S$ where the curve ends and beyond which it becomes complex valued but without horizons. \underline{Right}: Case $\ep=+1$, $\om < 0$, $\om < 2 \La_W$. The curves denote black membranes for $\La_W=1,0.5,-0.5, \om=-2 ~(\beta=-10)$ from left to right, respectively. There is a curvature singularity at $r_S$ where the curve ends and beyond which it becomes complex valued but without horizons.} \label{fig:Horava_horizon_k_zero_omega_positive_complex_left_and__omega_negative_plus_complex_left}
\end{figure}

For $\be<0$, there are no horizons and so the surface curvature singularity at
$r_S$ may be naked if it exists. In such case, the point-like singularity at
$r=0$ can be excluded since the allowed region in which the metric is real,
according to (\ref{lachota}), is $r>r_S$. The surface-like singularity exists
when $\om (2 \La_W-\om)<0$, {\it i.e.}, $\om > 0, \om > 2 \La_W$
(Fig. \ref{fig:Horava_horizon_k_zero_omega_positive_complex_left_and__omega_negative_plus_complex_left} (left))
or $\om < 0, \om < 2 \La_W$
(Fig.\ref{fig:Horava_horizon_k_zero_omega_positive_complex_left_and__omega_negative_plus_complex_left}
(right)). In these cases, the GR limit can be taken and according to
(\ref{GR_limit_3}) with $\be=4 \om M$, they run into the GR solutions with
$M<0, \La<0$ and $M>0, \La>0$, respectively, while $r_S\to 0$ matching the
naked curvature singularity of those GR solutions at the origin. For other
than these two cases, {\it i.e.,} $ 2\La_W < \om < 0$ or $0 < \om < 2 \La_W$,
there is neither real-valued metric for the whole region, nor the GR limit.
\subsubsection{Hyperbolic membrane solution $(k=-1)$}

The case $k=-1$ corresponds to the hyperbolic membrane. Its horizon structure can be understood as the intersections of the curves in Figs. \ref{fig:Horava_horizon_k_zero_omega_zero_AdSanddS} to
\ref{fig:Horava_horizon_k_zero_omega_positive_complex_left_and__omega_negative_plus_complex_left}
with an horizontal line at $f(r)=1$. Again, we classify the solutions according to the sign of $\beta$.

~

\noindent{\it A. Case $\beta>0$}:

~

\begin{figure}
\includegraphics[width=5.4cm,keepaspectratio]{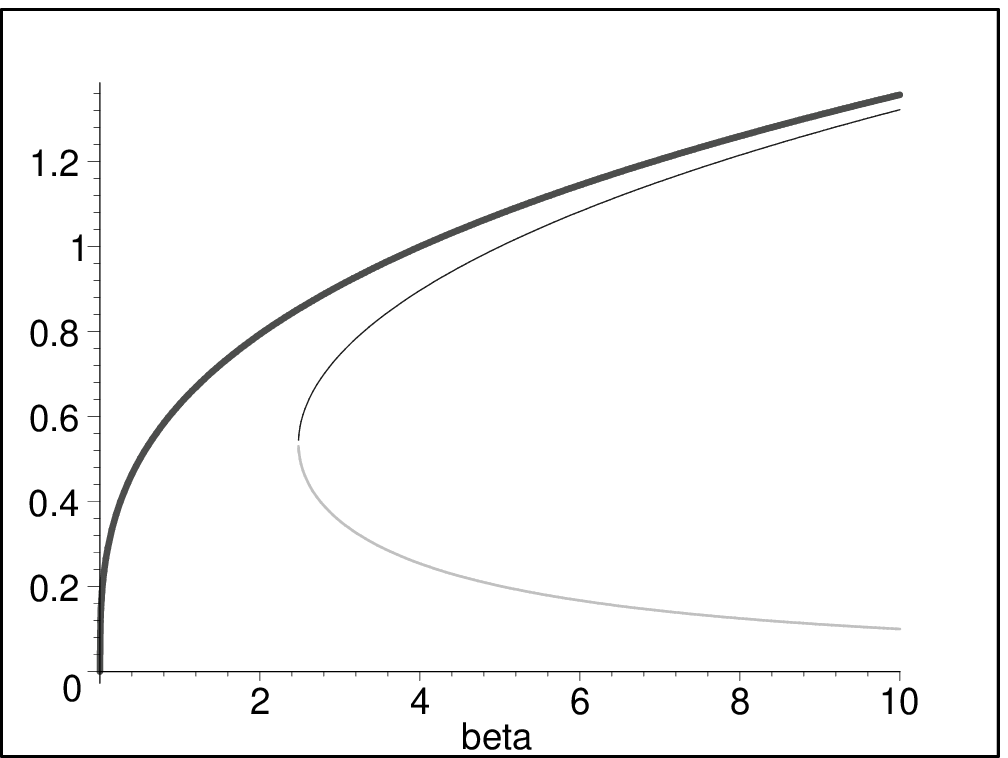}
%{Horizon_k_negative_omega_negative_plus_complex_right_201.eps}
\includegraphics[width=5.4cm,keepaspectratio]{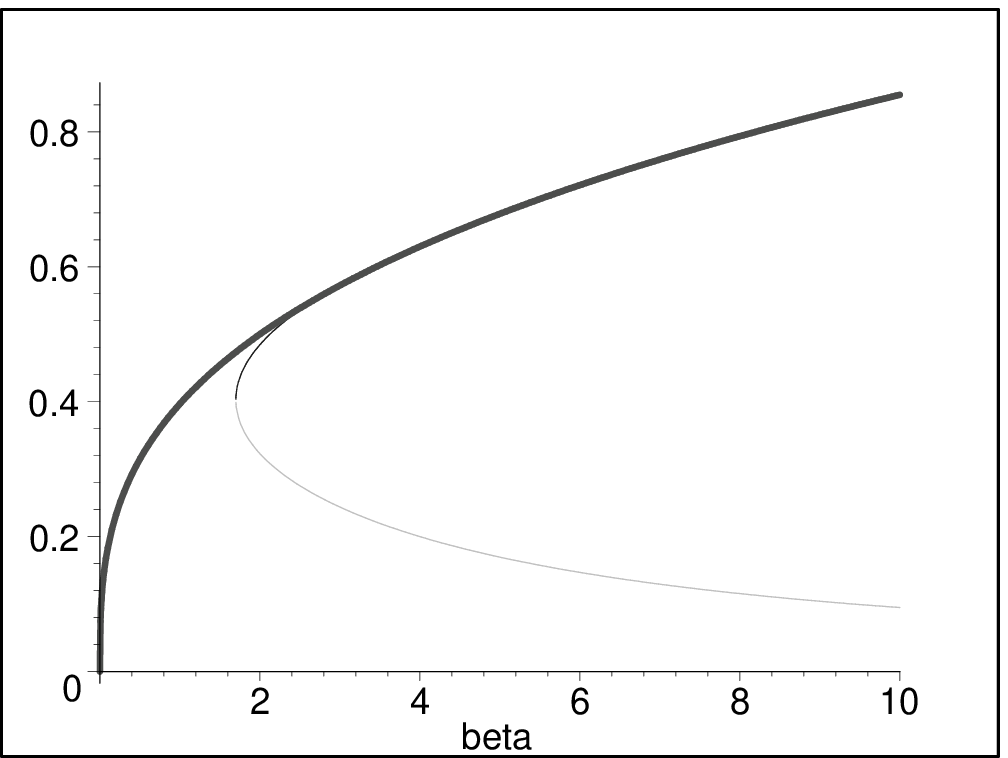}
%{Horizon_k_negative_omega_negative_plus_complex_right_101.eps}
\includegraphics[width=5.4cm,keepaspectratio]{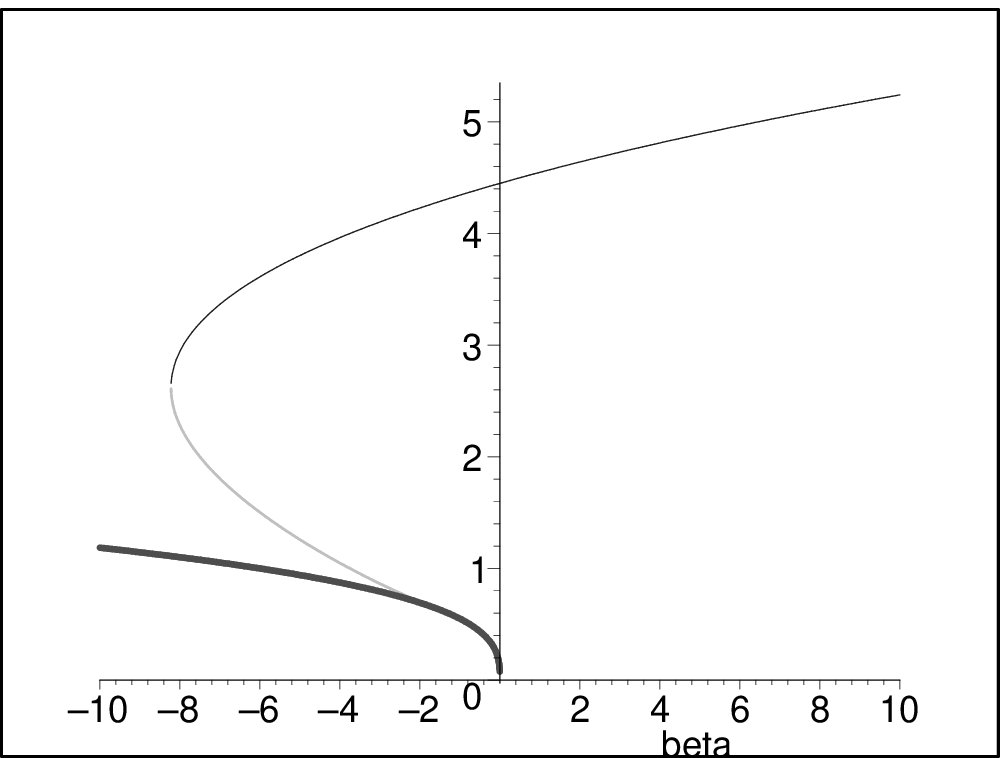}
%{Horizon_k_negative_omega_positive01.eps}
\caption{Plots of $r_{\pm}$ (two thin curves) and $r_S$ (thick curve)
vs. $\beta$, for $k=-1$. \underline{Left}: Case $\ep=+1,
2 \La_W <\om \leq \La_W <0$, in particular for $\La_W=-2, \om=-2$.
The plots show that $r_+ < r_S$ always, independently on $\be$ so that
the surface singularity is hidden whenever the horizon exists.
\underline{Center}: Case $\ep=+1, 2 \La_W <\La_W<\om <0$, in particular
for $\La_W=-5, \om=-2$. The plots show that $r_+ \leq r_S$, if the cosmological $r_+$ exists, for $\be \leq \tilde{\be}$ {($r_S=r_+\equiv\tilde{r}_+$ for $\be=\tilde{\be}$)} but for $\be >\tilde{\be}$ there is no cosmological horizon so that the surface singularity is naked. \underline{Right}: Case $\ep=-1, \om>0, \om>2 \La_W$, in particular for $\La_W=-0.5, \om=2$ (we have basically the same features for $k=``+1 ", \ep=+1, \om<0, \om<2 \La_W$).  The plots show that the curvature singularity $r_S$ is always inside the inner horizon $r_-$.} \label{fig:Horizon_k_negative_omega_negative_plus_complex_right_1_and__right_2_and_omega_positive}
\end{figure}

In this case, there is no inner horizon for the AdS branch ($\ep=-1$)
\footnote{In this case, the root formula for $r_-$ in (\ref{root}) does not
apply since it represents the horizon for the {\it un-physical} branch in
which the GR result (\ref{f_GR_limit}) is not recovered.}, but otherwise
the situation is more or less the same as the $k=0$ case. According to the
same classification as before, we have the following viable cases.
\begin{enumerate}[(a)]
\item $\om=0$ : Here, the surface-like singularity at $r=r_S$ does not
exist and there is no inner horizon for the AdS branch ($\epsilon=-1$).
The point-like singularity at the origin is hidden by a black membrane
horizon at $r_+$ for $\Lambda_W<0$, with the observer region $r>r_+$.
On the other hand, for the dS branch ($\epsilon=+1$), there is a black
membrane horizon at $r_-$ and a cosmological horizon at $r_+$ with the
observer region $r_-<r<r_+$ for $\Lambda_W>0, \beta>\beta_*$. The two
horizons meet at {$r_+=r_-\equiv r_*$ when} $\be=\be_*$. The case
$\La_W \leq 0$ can also provide an observer region $r_-<r$ with the black membrane horizon $r_-$ only (Fig. \ref{fig:Horava_horizon_k_zero_omega_zero_AdSanddS}).

\item $\om=2 \La_W$ :  This case corresponds to $``-\La_W " \ra ``\La_W"$ in the result of (a) and so all the properties can be understood just by flipping the sign of $\La_W$.

\item $\om > 0, \om > 2 \La_W, \epsilon=-1$ : Similar to the cases (a)
and (b), the surface-like curvature singularity at $r_S$ does not exist,
and the point-like singularity at $r=0$ is hidden by the outer black
membrane horizon at $r_+$, with the right signature of metric for the
observer region $r>r_+$.
(Fig. \ref{fig:Horava_horizon_k_zero_omega_positive_plus_and_omega_negative_plus}
(left)) This case flows to the case $M>0, \La<0$ of GR. For the case
$\La_W=0$, as in the flat membrane case, there is no observer region.

\item $\om < 0, \om < 2 \La_W, \epsilon=+1$ : Again, this case does not confront the surface-like curvature singularity since $r_S$ does not exist. The horizons at $r_\pm$ exist whenever $\be > \be_*$, the one at $r_+$ being a cosmological one, implying that the metric has the right signature in the observer region $r_-<r<r_+$. This case flows into the case $M<0, \La>0$ of GR (Fig. \ref{fig:Horava_horizon_k_zero_omega_positive_plus_and_omega_negative_plus} (right)). \\

Now, for the cases in which there is the surface-like curvature singularity at $r_S$, we can either have no horizon, or have a black membrane horizon at $r_+$ (case $2 \La_W > \om   >0$), but then since $r_+\leq r_S$ the surface-like singularity is naked as seen from our observer region (Fig. \ref{fig:Horava_horizon_k_zero_omega_positive_complex_and_omega_negative_plus_except} (left)), or have a cosmological horizon (case $2 \La_W < \om   <0$), and in such a case the surface-like singularity may be hidden depending on the value of the parameters (Fig. \ref{fig:Horava_horizon_k_zero_omega_positive_complex_and_omega_negative_plus_except} (right)). This leaves us with the following two viable cases.

\item $2 \La_W < \om  \leq \Lambda_W <0, \epsilon=+1$: Here, whenever $\be>\be_*$, there exists a cosmological horizon $r_+$ as well as a
black membrane horizon $r_-$
(Fig. \ref{fig:Horava_horizon_k_zero_omega_positive_complex_and_omega_negative_plus_except}
 (right), Fig. \ref{fig:Horizon_k_negative_omega_negative_plus_complex_right_1_and__right_2_and_omega_positive} (left)).
 The point-like singularity at $r=0$ is hidden by the black membrane
 horizon $r_-$ with the observer region $r_-<r<r_+$. The curvature singularity at $r_S$ {is} always beyond the cosmological horizon $r_+ \le r_S$.

\item $2 \La_W <\Lambda_W < \om  <0, \epsilon=+1$: In this case
there exists a black membrane horizon $r_+$ when $\be>\be_*$ so that the
point-like singularity at $r=0$ is also hidden, as in the case (e).
However, in contrast to that case the surface-like singularity can be
hidden only for $\be \leq \tilde{\be}$, where $\tilde{\be}$ is defined
as the value of $\beta$ for which there is a merging of the cosmological horizon
with the surface-like singularity, $r_S=r_+ \equiv\tilde{r}_+$ for $\be=\tilde{\be}$
(Fig. \ref{fig:Horava_horizon_k_zero_omega_positive_complex_and_omega_negative_plus_except} (right),
Fig.  \ref{fig:Horizon_k_negative_omega_negative_plus_complex_right_1_and__right_2_and_omega_positive} (center)).
For $\be>\tilde{\be}$, there is no cosmological horizon at $r_+$ behind
which the surface-like singularity would be hidden; this can be seen in
the absence of the larger root for $f(r)=0$
(Fig. \ref{fig:Horava_horizon_k_zero_omega_positive_complex_and_omega_negative_plus_except} (right),
top curve) \footnote{In this case again, the root formula for $r_+$ in
(\ref{root}) does not apply due to the same reason as in the footnote 12.}. So, this is the case in which the surface-like singularity can penetrate to our observer region unless $\be$ is constrained as $\be \leq \tilde{\be}$
\footnote{This phenomena may be interpreted as the horizons being
{\it melted away} or {\it eaten} by the surface-like singularity since the latter carries infinite temperature as can be seen in Sec. IV.}.
\end{enumerate}
~

\noindent{\it B. Case $\beta<0$}:

~

The situation is quite different for the $\beta<0$ case, where viable black membrane solutions exist in the following exceptional case

\begin{enumerate}[(c')]
\item $\om>0, \om > 2\La_W, \epsilon=-1$ : This is the case where $\be$ can be negative so that we have a chance to satisfy $0> \be >\be_*$ with
the inner and outer black hole horizon at $r_{\pm}$ (Fig. \ref{fig:beta_omega_negative_and_positive} (left),
Fig. \ref{fig:Horava_horizon_k_negative_beta_negative} (left)).

Here, it is important to note that even if the surface-like curvature singularity at $r_S$ is present, it sits always inside the black membrane horizon, {\it i.e.,} $r_+ >r_S$. Actually, as we increase $\beta$ from its minimum value $\be_*<0$, the inner horizon $r_-$ shrinks and meets the
curvature singularity $r_S$ at $r=\tilde{r}_-$, $\be=\tilde{\be}$
(Fig. \ref{fig:Horizon_k_negative_omega_negative_plus_complex_right_1_and__right_2_and_omega_positive} (right)).
On the other hand, for $\be >  \tilde{\be}$, there is no inner horizon,
but the curvature singularity at $r_S$ is not naked since the outer black membrane horizon is always outside the singularity, {\it i.e.,}
$r_+ > r_S$. For $\be < \be_*$, the horizons are not formed so that the surface singularity is naked. This case flows to the exceptional solution of $M<0, \La<0$ in GR.
\end{enumerate}

\subsubsection{Spherical membrane solution $(k=+1)$}

Finally, the spherical membrane case $k=+1$ is known as the black hole solution.  Its horizon structure can be understood as the intersections of the curves in Figs.
\ref{fig:Horava_horizon_k_zero_omega_zero_AdSanddS} to
\ref{fig:Horava_horizon_k_zero_omega_positive_complex_left_and__omega_negative_plus_complex_left}
with an horizontal line at $f(r)=-1$. Similarly to the previous
cases, we classify the solutions according to the sign of $\beta$.

~

\noindent{\it A. Case $\beta>0$}:

~

According to the same classification as before, we have the following viable cases.

\begin{enumerate}[(a)]
\item $\om=0$ : Here, the surface-like singularity does not exist. The point-like singularity at $r=0$ is hidden by an inner black hole horizon $r_-$ as well as by an outer black hole horizon $r_+$, with the correct signature of the metric for $r>r_+$ whenever $\be>\be_*$ in the AdS branch ($\epsilon=-1$)
(Fig. \ref{fig:Horava_horizon_k_zero_omega_zero_AdSanddS} (left)).
Due to the absence of an inner horizon, in the dS branch ($\epsilon=+1$), there are no viable solutions; the singularity at the origin is always naked as seen from the observer region $0<r<r_+$ with a cosmological horizon $r_+$ (Fig. \ref{fig:Horava_horizon_k_zero_omega_zero_AdSanddS} (right)).

\item $\om=2 \La_W$ : This case corresponds to $-\La_W  \ra \La_W$ in the result of (a) and so all the properties can be understood just by flipping the sign of $\La_W$.

\item $\om > 0, \om > 2 \La_W, \ep=-1$ : Similar to the cases (a) and (b), the surface-like curvature singularity does not exist, and the point-like singularity at $r=0$ is hidden by {an inner black hole} horizon at $r_-$ as well as by an outer black hole horizon at $r_+$ whenever $\be>\be_*$, with the right signature of metric for the spatial coordinate $r>r_+$ (Fig. \ref{fig:Horava_horizon_k_zero_omega_positive_plus_and_omega_negative_plus} (left)). This case flows to the case $M>0, \La<0$ of GR.

\end{enumerate}

Notice that the case $\om < 0, \om < 2 \La_W, \ep=+1$ is not viable for $k=+1$, since there is only a cosmological horizon $r_+$ and so the singularity at $r=0$ is never hidden in the observer region $0 \leq r < r_+$
(Fig. \ref{fig:Horava_horizon_k_zero_omega_positive_plus_and_omega_negative_plus} (right)). On the other hand, for the cases in which there is a surface-like curvature singularity at $r_S$, we can either have no horizon, or have a cosmological horizon at $r_+$ (case $2 \La_W < \om   < 0, \epsilon=+1$), but then the point-like singularity {at $r=0$} is naked as seen from our observer region
(Fig \ref{fig:Horava_horizon_k_zero_omega_positive_complex_and_omega_negative_plus_except} (right)); or have a black hole horizon (case $2 \La_W > \om   >0, \epsilon=-1$), and in such case the surface-like singularity is naked (Fig. \ref{fig:Horava_horizon_k_zero_omega_positive_complex_and_omega_negative_plus_except} (left)). This leaves us with no viable cases.

~

\noindent{\it B. Case $\beta<0$}:

~

\begin{figure}
\includegraphics[width=7.3cm,keepaspectratio]{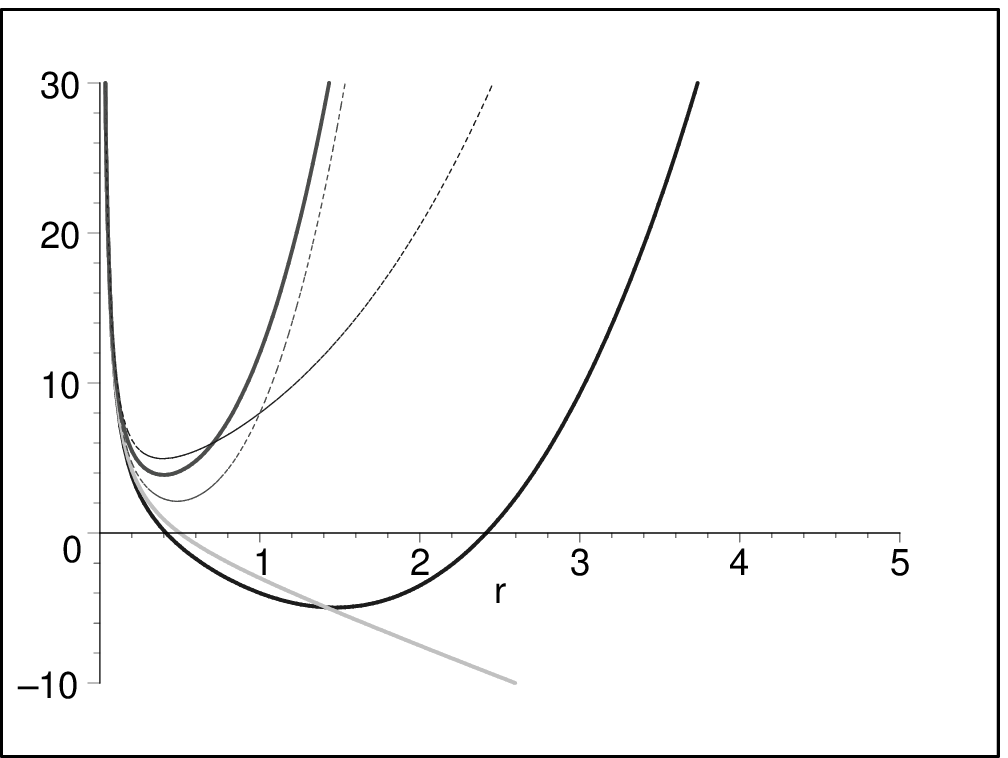}
%{beta01.eps}
\qquad\qquad
\includegraphics[width=7.3cm,keepaspectratio]{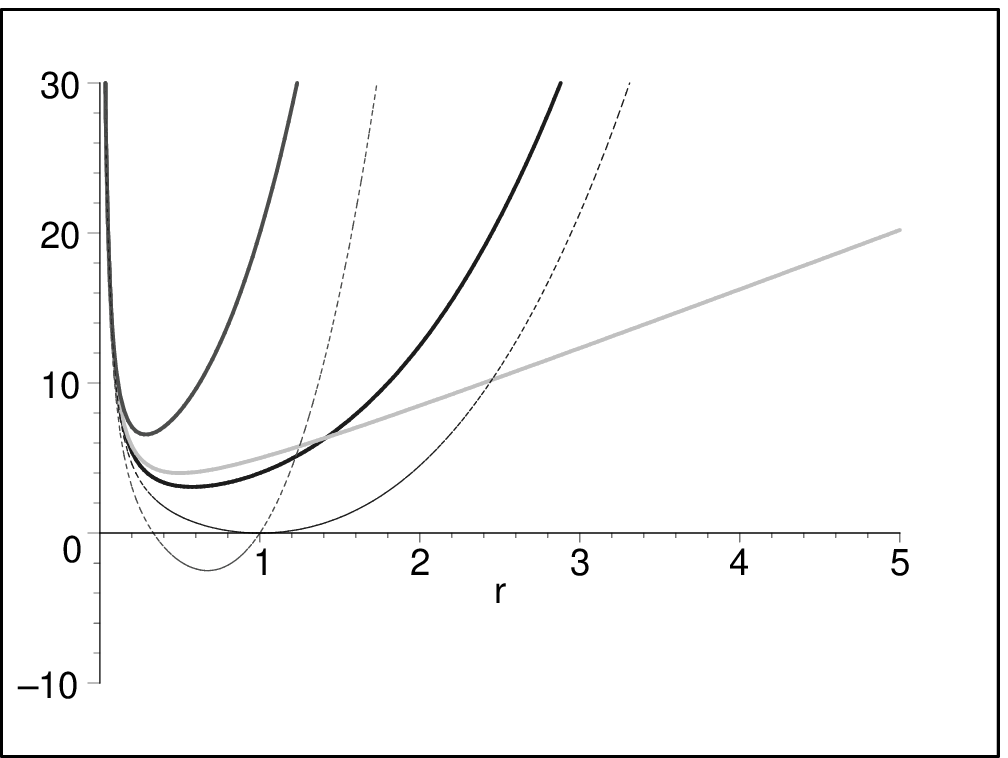}
%{beta_omega_negative01.eps}
\caption{Plots of $\beta$ vs. $r_+$. The thick curves denote the hyperbolic membrane $(k=-1)$ for $\La_W=-1,0,3$ (bottom to top, from left) and thin dotted curves denote the spherical membrane $(k=+1)$ for $\La_W=-1,3$ (top to bottom, from left). \underline{Left}: Case $\om>0$. For the $k=-1$ case, $\be$ can be negative so that we have a chance to satisfy the condition $\be_* < \be <0$ for the existence of horizons $r_{\pm}$ with $\be<0$. However, this is not possible for the $k=+1$ case. The plots are for $\om=2$. \underline{Right}: Case $\om<0$. Here, the situation is the opposite  and for the $k=+1$ case, $\be$ can be negative so that we have a chance to satisfy the condition $\be_* <\be <0$ for the existence of horizons $r_{\pm}$ with $\be<0$. However, this is not possible for the $k=-1$ case. The plots are for $\om=-2$.} \label{fig:beta_omega_negative_and_positive}
\end{figure}

\begin{figure}
\includegraphics[width=7.3cm,keepaspectratio]{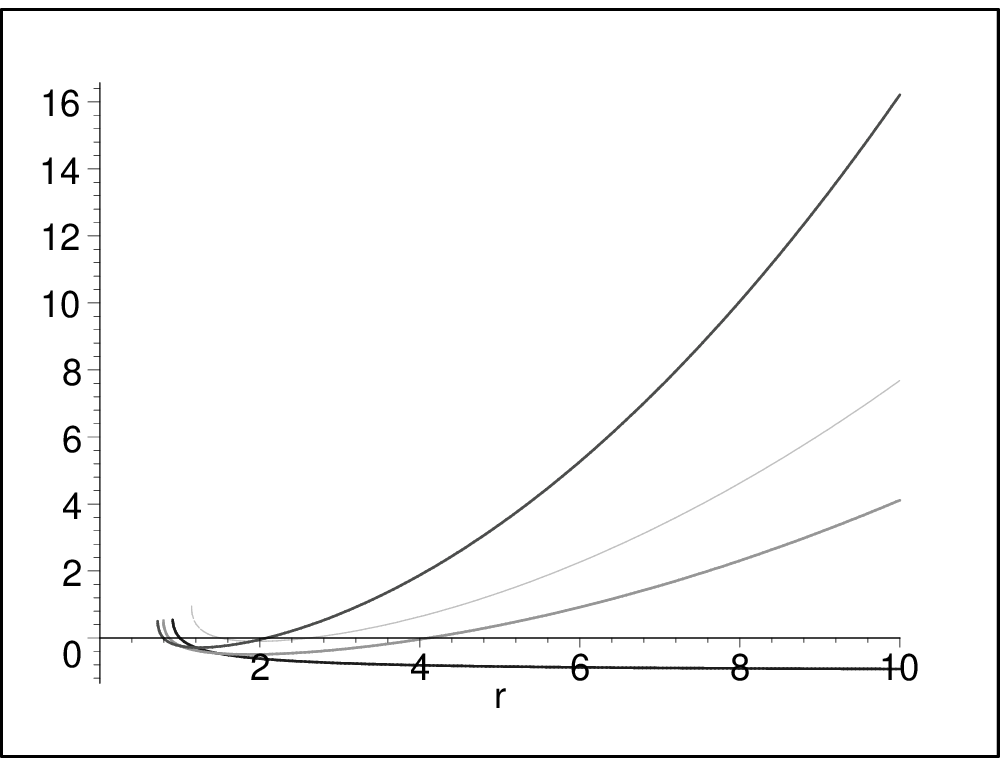}
%{k_negative_beta_negative01.eps}
\qquad\qquad
\includegraphics[width=7.3cm,keepaspectratio]{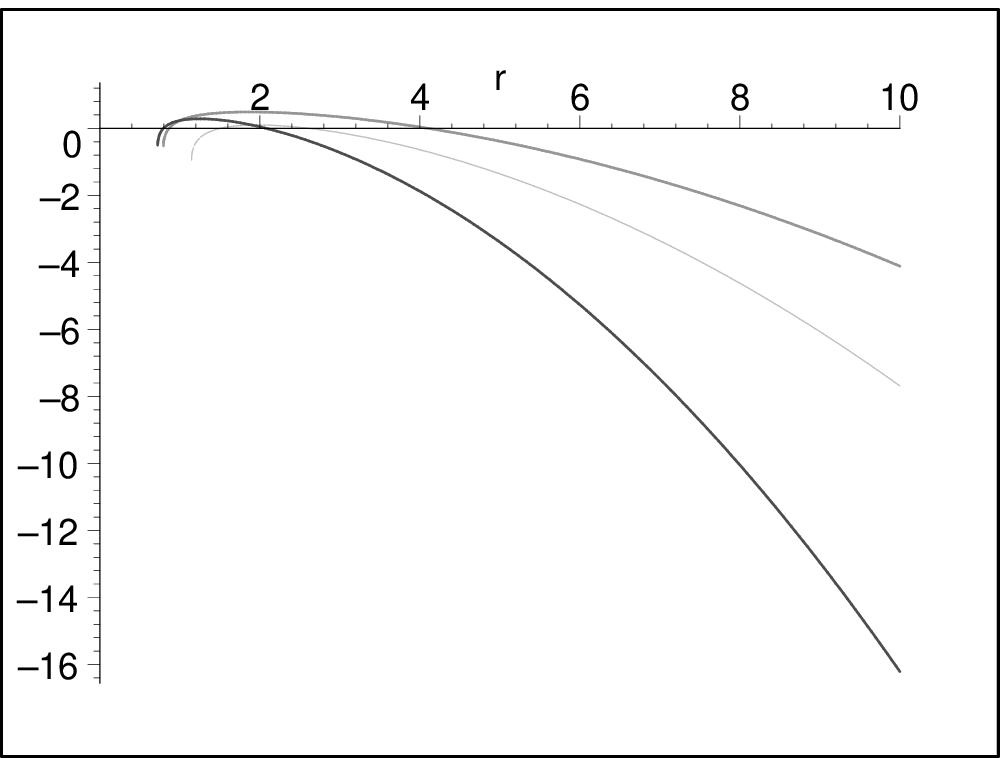}
%{k_negative_beta_negative_plus01.eps}
\caption{Plots of $f(r)$ for $\beta<0$, $\om (2 \La_W-\om)<0$.
\underline{Left}: Case $k=-1, \ep=-1, \om > 0, \om > 2 \La_W$. The curves denote black hyperbolic branes with the inner and outer horizons for $\La_W=-1, -0.5, 0, 0.5, \om=2 ~(\be=-3)$ from left to right, respectively. \underline{Right}: Case $k=+1, \ep=+1, \om < 0, \om < 2 \La_W$. The curves denote black membranes with the inner and outer horizons for $\La_W=1, 0.5, -0.5, \om=-2 ~(\be=-3)$ from left to right, respectively.}
\label{fig:Horava_horizon_k_negative_beta_negative}
\end{figure}

For $\beta<0$ in the AdS branch ($\ep=-1$), we have the same situation
as in the $k=0$ case, having no horizon and no viable solutions without naked singularity (Fig. \ref{fig:Horava_horizon_k_zero_omega_positive_complex_left_and__omega_negative_plus_complex_left} (left)). In the dS branch ($\ep=+1$), there is a curvature singularity $r_S$ at the boundary of the real metric with $\om < 0, \om < 2 \La_W$ and otherwise, there is no real-valued metric for the whole region as in the $k=0$ case. This
leaves us with the only interesting one as follows.
\begin{enumerate}[(d')]
 \item $\om < 0, \om < 2 \La_W, \ep=+1$ : In this case, an inner black
 hole horizon at $r_-$ and  an outer cosmological horizon at $r_+$ exist, as long as $\beta>\beta_*$ {(Fig. \ref{fig:beta_omega_negative_and_positive},
 Fig. \ref{fig:Horava_horizon_k_negative_beta_negative} (right))}, and
 the observer region is $r_-<r<r_+$. This case flows to the case $M>0, \La>0$ of GR. As we increase $\beta$ from its minimum value $\be_*<0$,
the black hole horizon $r_-$ shrinks and meets the curvature
singularity $r_S$ at $r=\tilde{r}_-$, $\be=\tilde{\be}$
(Fig. \ref{fig:Horizon_k_negative_omega_negative_plus_complex_right_1_and__right_2_and_omega_positive} (right)).
On the other hand, for $\be >  \tilde{\be}$, there is no black hole horizon
so that the curvature singularity at $r_S$ becomes naked. This situation is analogous to that of $k=-1$ in Fig.
\ref{fig:Horizon_k_negative_omega_negative_plus_complex_right_1_and__right_2_and_omega_positive} (center), but now the surface-like singularity can penetrate to our observer region from {\it inside} the black hole horizon unless $\be$ is
constrained as $\be \leq \tilde{\be}$. This achieves a close analogy with the singularity at $r=0$ in GR case, which can be naked unless $M$ is constrained as $M>0$ for $\La>0$.
\end{enumerate}


To conclude this section, we have classified all the viable $(\lambda=1)$ static black membrane solutions without naked curvature singularities as seen from the observer region, where the metric has the right signature for Ho\v{r}ava gravity. The solutions are classified by $\om, \La_W$, and $\be$, and we have found several interesting black membrane solutions which do not exist in GR. In particular, we have found that there are black plane ($k=0$) and hyperbolic ($k=-1$) branes even in the dS branch (case (d), (e) for the former and case (d), (e), (f) for the latter), where there is a cosmological horizon, as well as in the AdS branch. This implies that, in these particular cases, some additional ``attraction" is generated due to the higher-derivative effects of Ho\v{r}ava gravity so that the membranes can be formed by overcoming the global repulsion in the dS branch.

\section{Thermodynamics}

For the AdS branch, the solution (\ref{solution}) has two horizons
generically and the Hawking temperature for the outer horizon $r_+$ is given
by \footnote{Due to the lack of Lorentz invariance in UV, the very meaning
of the horizons and Hawking temperature would be changed from the
conventional ones. The light cones would differ for different wavelengths
and so different particles with different dispersion relations would see
different Hawking temperature and entropies, and the Hawking spectrum would
not be thermal. But from the recovered Lorentz invariance in the IR
(with $\la=1$), the usual meaning of the horizons and $T_H$ as the
Hawking temperature would be ``emerged" for long wavelengths. The calculation
and meaning of the temperature should be understood in this context. }
\begin{\eq}
T_H=\f{3 \La_W^2 r_+^4 +2 k (\om-\La_W)r_+^2 -k^2 }{ 8 \pi r_+ (k
+(\om-\La_W)r_+^2)}\, . \label{temp}
\end{\eq}
\begin{figure}
\includegraphics[width=7.3cm,keepaspectratio]{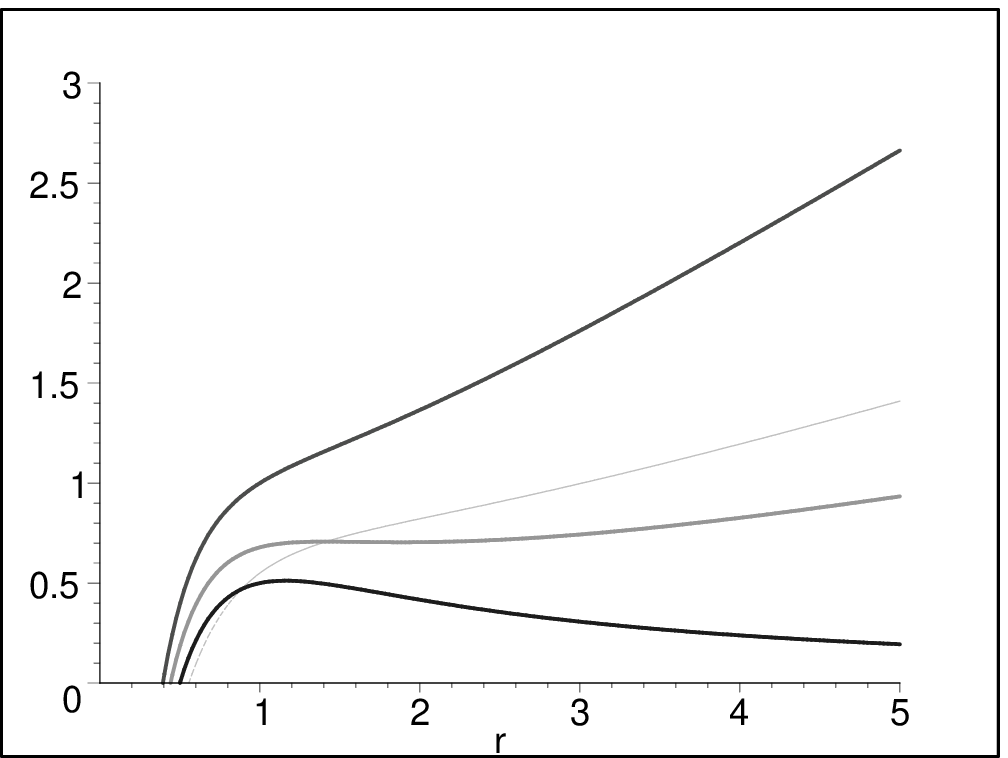}
%{temp_k_positive_omega_positive01.eps}
\qquad\qquad
\includegraphics[width=7.3cm,keepaspectratio]{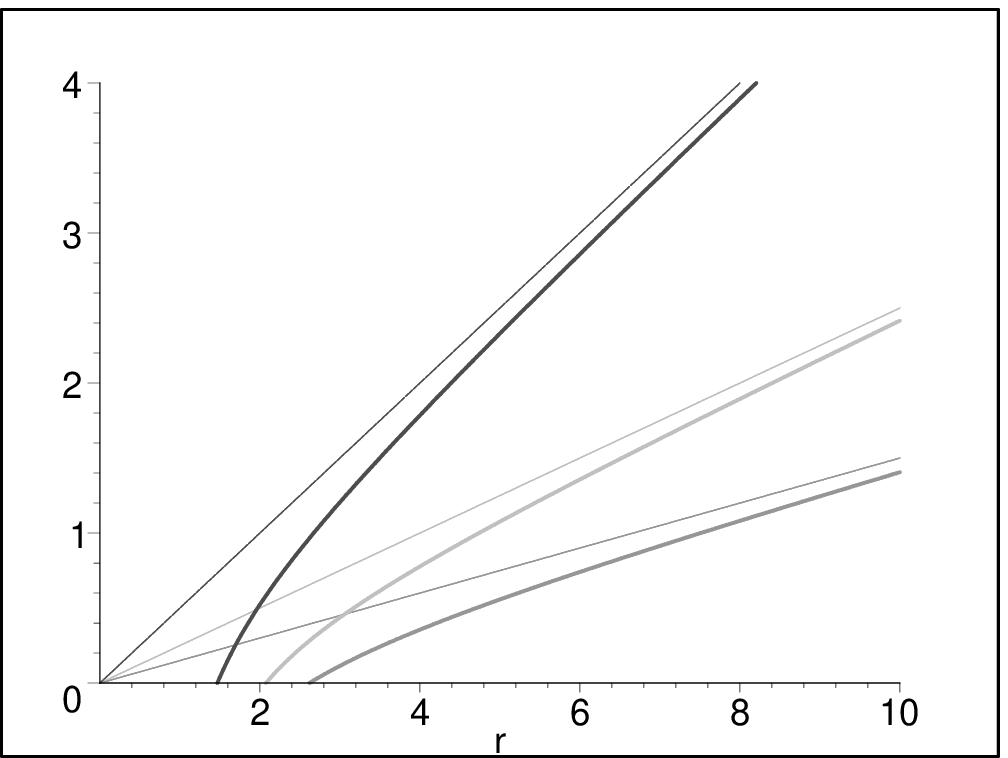}
%{temp_k_zero_negative_omega_positive01.eps}
\caption{Plots of $4 \pi T$ vs. the outer horizon radius $r_+$ for the AdS branch with $\om>0, \om>2 \La_W, \ep=-1$. \underline{Left}: Black hole case ($k=+1$), in particular for $\La_W=-1,-0.5,0$ (top to bottom, thick curves),
{$0.5$ (thin curve), $\om=2$}. \underline{Right}: Hyperbolic membrane case ($k=-1$) (thick curves), and flat membrane case ($k=0$) (thin curves), in particular for $\La_W=-1$ (top), $-0.5$ (bottom), $0.5$ (middle), $\om=2$.}
\label{fig:temp_k_positive_omega_positive_AdS}
\end{figure}

Note that this temperature diverges when the horizon radius $r_+$ coincides with the point-like singularity at $r=0$ and, interestingly, also when it coincides with the surface-like singularity at $r_S$, where the denominator vanishes.

Fig. \ref{fig:temp_k_positive_omega_positive_AdS} (left) shows that the black
hole temperature interpolates between the asymptotically AdS cases
(above three curves) and flat (bottom curve) case. There exists an extremal black hole limit of vanishing temperature where the inner horizon $r_-$ meets with the outer horizon $r_+$ at $r=r_*$ and the integration constant $\be$ gets its minimum $\be_*$. For smaller black holes of $r_+<r_*$, the black hole temperature becomes negative, implying a thermodynamics instability. This may provide the minimum size for a thermodynamically stable black hole. The flat ($k=0$) and hyperbolic ($k=-1$) membranes have the same properties
(Fig. \ref{fig:temp_k_positive_omega_positive_AdS} (right)), even though we have zero minimum radius for $k=0$.

\begin{figure}
\includegraphics[width=5.4cm,keepaspectratio]{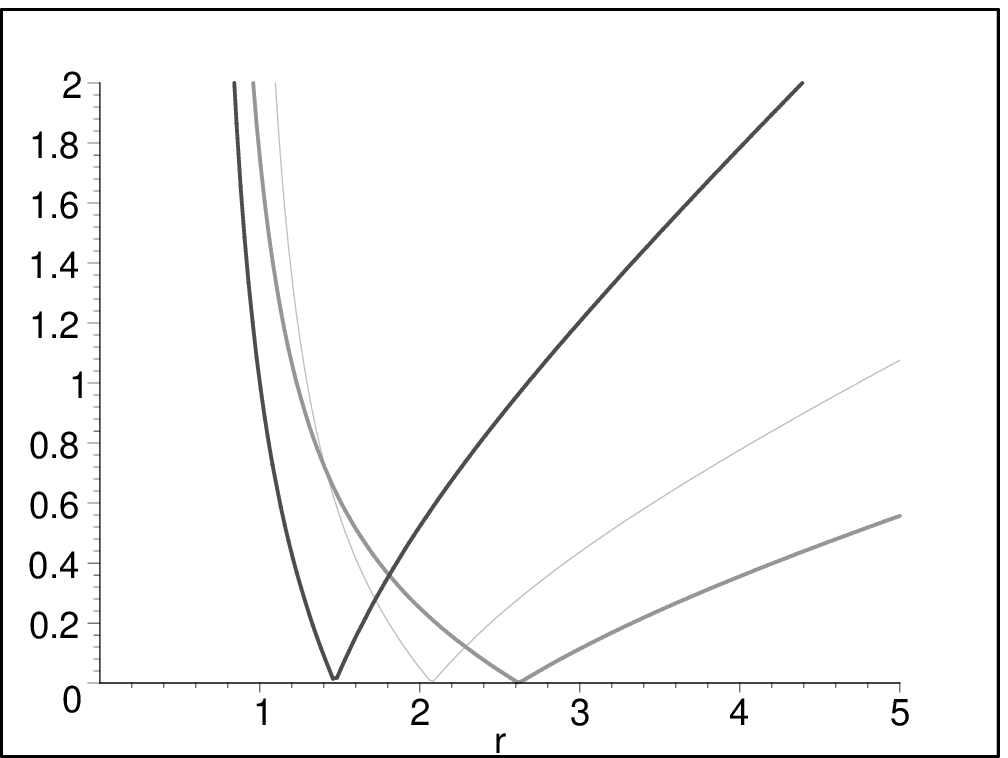}
%{temp_k_positive_omega_negative_complex_left01.eps}
\includegraphics[width=5.4cm,keepaspectratio]{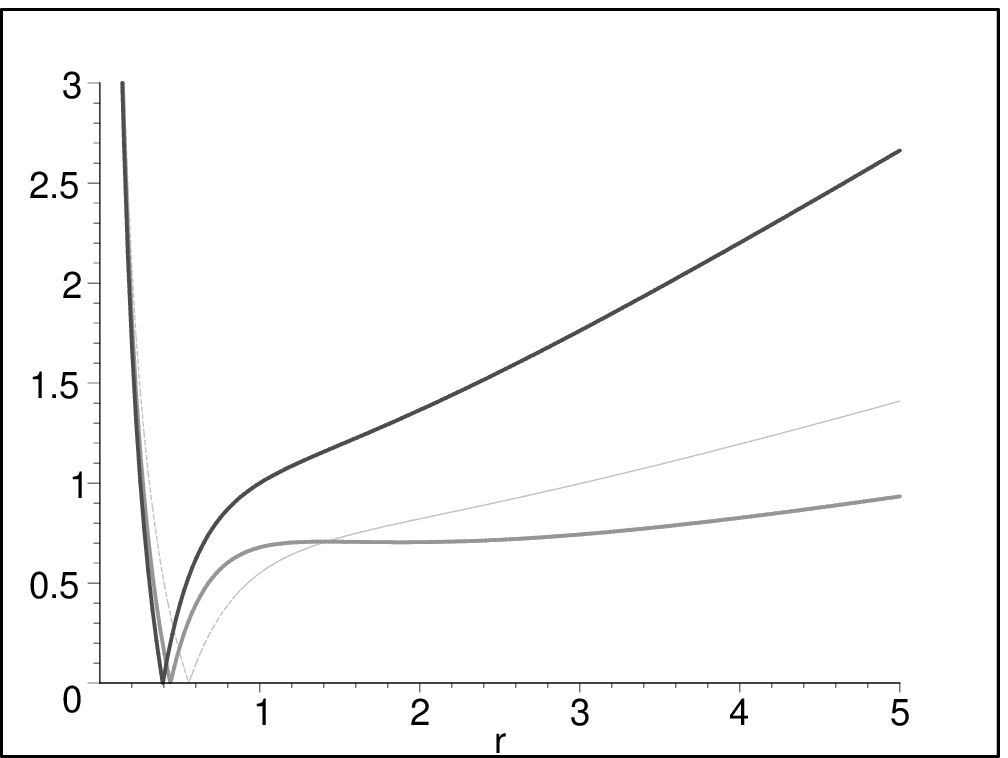}
%{temp_k_negative_omega_negative01.eps}
\includegraphics[width=5.4cm,keepaspectratio]{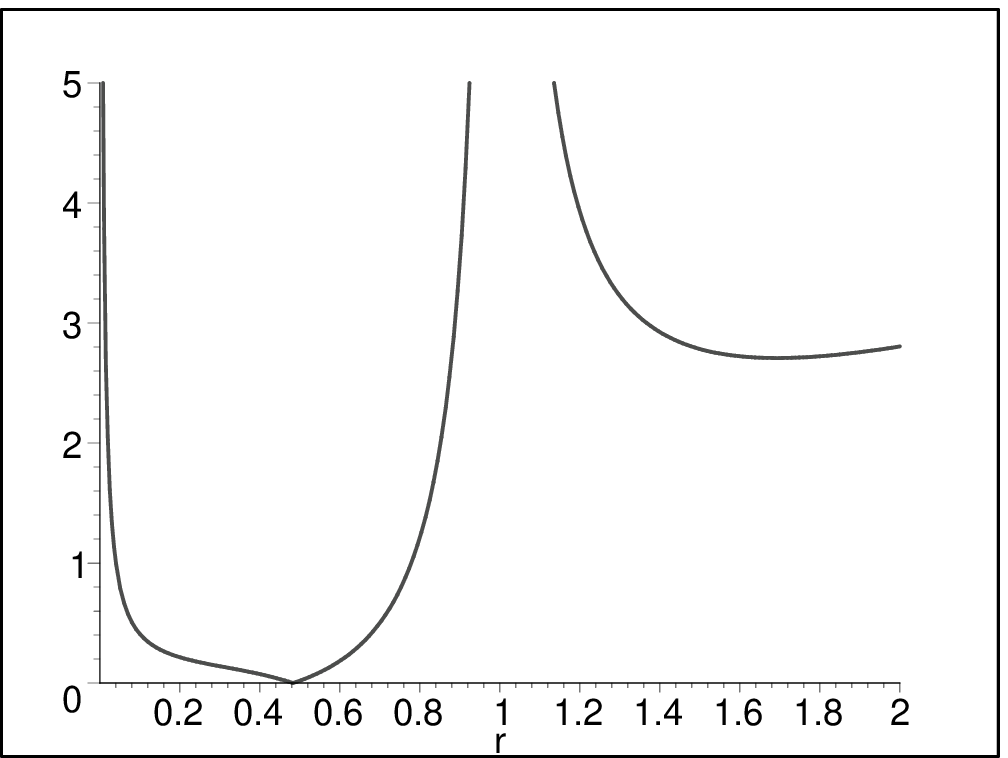}
%{temp_k_negative_omega_negative_complex_right01.eps}
\caption{Plots of $4 \pi T$ vs. black hole or membrane's horizon radius
$r_-$ (left part) and cosmological horizon radius $r_+$ (right part) for
the dS branch ($\ep=+1$). \underline{Left}: Black hole case $(k=+1)$
with $\om<0, \om<2 \La_W$, in particular for $\La_W=1,0.5$
(top to bottom, thick curves),
$\La_W=-0.5$ (thin curve), $\om=-2$. The black hole temperature becomes infinity at $\tilde{r}_-$ where the black hole's horizon coincides with the surface-like singularity. \underline{Center}: Hyperbolic membrane case $(k=-1)$ with $\om<0, \om< 2 \La_W$, in particular for $\La_W=1,0.5
$ (top to bottom, thick curves), $\La_W=-0.5$ (thin curve), $\om=-2$.
\underline{Right}: Hyperbolic membrane case with $ \La_W<\om<0$, in particular for $\La_W=-3, \om=-2$. The cosmological horizon temperature becomes infinity at $\tilde{r}_+=1$, where the cosmological horizon coincides with the surface-like singularity.
}
\label{fig:temp_k_negative_omega_negative_complex_right_dS}
\end{figure}

For the dS branch, the solution (\ref{solution}) can have the inner black hole/membrane horizon at $r_{-}$ and the cosmological horizon at $r_+$, and the temperature for the black hole/membrane is given by (\ref{temp}) also but
now with $r_-$ in place of $r_+$. There is an extremal limit of vanishing
temperature at $r_+=r_-= r_*$ in which the black hole/membrane horizon
coincides with the cosmological horizon, {\it i.e.,} the Nariai limit
(Fig. \ref{fig:temp_k_negative_omega_negative_complex_right_dS}). We see that the temperature becomes infinity at the vanishing limit of membrane radius where the black membrane horizon at $r_-$ meets with the point-like singularity at $r=0$, as in the case of \Sch or Schwarzschild-de Sitter black hole in GR
(Fig. \ref{fig:temp_k_negative_omega_negative_complex_right_dS} (center)). It is interesting to note that there is another (positive) infinite temperature point at $r=\tilde{r}_{\pm}$ where the black membrane horizon at $r_-$ or the cosmological horizon at $r_+$ coincides with the surface-like singularity at $r_S$, which provides lower/upper bounds for the black hole/cosmological horizons radius (Fig. \ref{fig:temp_k_negative_omega_negative_complex_right_dS} (left) and (right)). So, the occurrence of the infinite temperature would be a
reflection of the coincidence of a curvature singularity with the
black hole/membrane or cosmological horizons \footnote{This seems to be quite
generic behaviors when the Killing and apparent horizons coincide. But, this
does not be seems to be true otherwise. See for example \ci{Park:2013}.}.

Finally, we note that one can also consider the first law of black membrane
thermodynamics as in the usual form, for the black membrane horizon $r_+$
\begin{\eq}
d{\cal M}=T_H d {\cal S}\, ,
\end{\eq}
with the black membrane's mass and entropy
\begin{\eq}
{\cal M}&=&\f{\kappa^2 \mu^2 \Om_k \be}{16}\, , \no \\
{\cal S}&=&\f{\pi \kappa^2 \mu^2 \Om_k}{4} \left( (\om-\La_W) r_+^2 + 2 k ~\mbox{ln} r_+\right) +{\cal S}_0\, ,
\label{entropy}
\end{\eq}
respectively, up to an arbitrary constant ${\cal S}_0$ \ci{Myun,Cai:0910}.
However, as far as we know, the very meaning of the entropy in Ho\v{r}ava gravity is not quite clear and not well established yet \ci{Cai}.

\section{Connection to time-dependent cosmological solutions}

So far, we have studied the viable solutions, without naked singularities,
of black holes and black membranes with dS or AdS asymptotics, for $\la=1$,
which matches with GR in the IR. In GR, there is a close connection between
a static metric and a time-dependent cosmological solution via coordinate
transformations which mix space and time. For example, the $dS_4$ metric in
static coordinates can be mapped into a flat FLRW metric in planar
coordinates \cite{Kim:2002}. Since Einstein equations are invariant under
such change of coordinates, the static solutions are mapped into cosmological
solutions. However, in Ho\v{r}ava gravity, this correspondence does not hold anymore, due to lack of full diffeomorphism invariance, and we cannot get a direct connection between those two spacetimes. In this section, we study whether some information, in particular the conditions for the viable solutions without naked singularities, can be mapped from static black holes or
membranes to cosmology solutions, even in the absence of a direct connection.
To this end, we consider a homogeneous and isotropic cosmological ansatz for
the action (\ref{horava}) with the standard FLRW form
\begin{\eq}
ds^2=-c^2
dt^2+a^2(t)\left(\frac{dr^2}{1-\textbf{k} r^2
}+r^2\left(d\theta^2+\sin^2\theta
d\phi^2\right)\right)\, ,
\label{FRW}
\end{\eq}
where the three-dimensional spatial curvature $\textbf{k}=+1,0,-1$ correspond
to a closed, flat, and open universe, respectively. The curvature invariants
of the metric (\ref{FRW}) are given by
\begin{\eq}
R^{(3)}=\f{6 \textbf{k}}{a^2}\,,\qquad \qquad~{K^i}_i=\f{3 \dot{a}}{a}\,
\end{\eq}
and we see that there is only an initial curvature singularity at $a(t)=0$.

Assuming the matter contribution to be of the form of a perfect fluid with the energy density $\rho$ and pressure $p$, we find that

\begin{\eq}
\left(\f{\dot{a}}{a}\right)^2&=&\frac{\kappa^2}{6(3 \lambda-1)}
\left[\rho
- \frac{3\kappa^2\mu^2}{8(3 \lambda-1)} \left( \f{\textbf{k}^2}{a^4}+ \f{2 \textbf{k} (\om-\La_W )}{a^2}+ \La_W^2 \right) \right]\, ,  \label{F1}
\\
\f{\ddot{a}}{a}&=&\frac{\kappa^2}{6(3 \lambda-1)} \left[-\f{1}{2}(\rho+3 p) +\frac{3 \kappa^2\mu^2}{8(3 \lambda-1)} \left( \f{
\textbf{k}^2}{a^4}- \La_W^2 \right) \right]\, . \label{F2}
\end{\eq}
Note that the $1/a^4$ term, which is the contribution from the
higher-derivative terms in the action (\ref{horava}), exists only
for $\textbf{k} \neq 0$ and becomes dominant for small $a(t)$, implying that
the cosmological solutions of GR are recovered at larger scales. As usual, the second equation has a first integral, whose value is completely fixed by the first, which turns out to be the only independent equation of the system. Here, we have not restricted to $\la=1$ like the previous sections since the following analysis is more generally valid for arbitrary values of $\la>1/3$, which would be quite useful in cosmology \ci{Keha,Park:0905,Park:0906}.

In order to study the solutions for the scale factor $a(t)$, it is useful to
consider the effective potential
\begin{\eq}
V_{\rm eff}=\frac{\kappa^2 }{32 (3\la -1)^2}
 \left[ -2 (3\la -1)
\rho  a^2+\kappa^2 \mu^2 \left( \f{{\bf k}^2}{a^2} + 2{\bf k} (\om-\Lambda_W) + \Lambda_W^2 a^2 \right) \right]\, ,
\end{\eq}
on the effective mechanical equation $\dot{a}^2/2+ V_{\rm eff}=0$ for a
particle of unit mass and zero energy. In this picture, a non-singular cosmology corresponds to a situation in which there are bouncing points that prevent the particle to reach the origin $a=0$. The bouncing points are located at the values of the scale factor at which $\dot a^2/2=-V_{\rm eff}=0$.

For the case of a flat ($\textbf{k}=0$) universe solution, there is no
contribution to the effective potential arising from the higher-derivative terms in Ho\v{r}ava gravity, so that we have basically the same situation as in GR where the initial singularity exists always \footnote{For the de Sitter-type universe solution, with a exponentially growing or decaying scale factor $a(t)$, the singularity, $a(t)=0$, is pushed to the infinite past
$t \rightarrow -\infty$ or the infinite future $t \rightarrow +\infty$,
respectively but the Big Bang or Big Crunch singularity problem remains still.}
unless we introduce some  exotic matter that violates energy conditions, {\it i.e.,} $\rho <0$.

However, for the non-flat $({\bf k} \neq 0)$ cases, non-singular vacuum cosmology solutions can exist \ci{Park:0905} \footnote{For the non-singular cosmology solutions in the presence of matter, see \ci{Wang, Mina, Maed}. } if the following conditions,
\begin{\eq}
\omega (\om- 2 \La_W) \geq 0\,,
%\qquad\qquad
~~{\bf k} (\om-\La_W)<0\,
\label{nonsingular_FRW}
\end{\eq}
are satisfied. In such a case, the bouncing points for $V_{\rm eff}=0$
exist, at the values of the scale factor given by
\begin{\eq}
(a^{\pm})^2=\f{-{\bf k} (\om-\La_W) \pm \sqrt{{\bf k}^2 \om (\om -2 \La_W)}}{\La_W^2}\, .
\label{chonga}
\end{\eq}

\begin{figure}
\includegraphics[width=7.3cm,keepaspectratio]{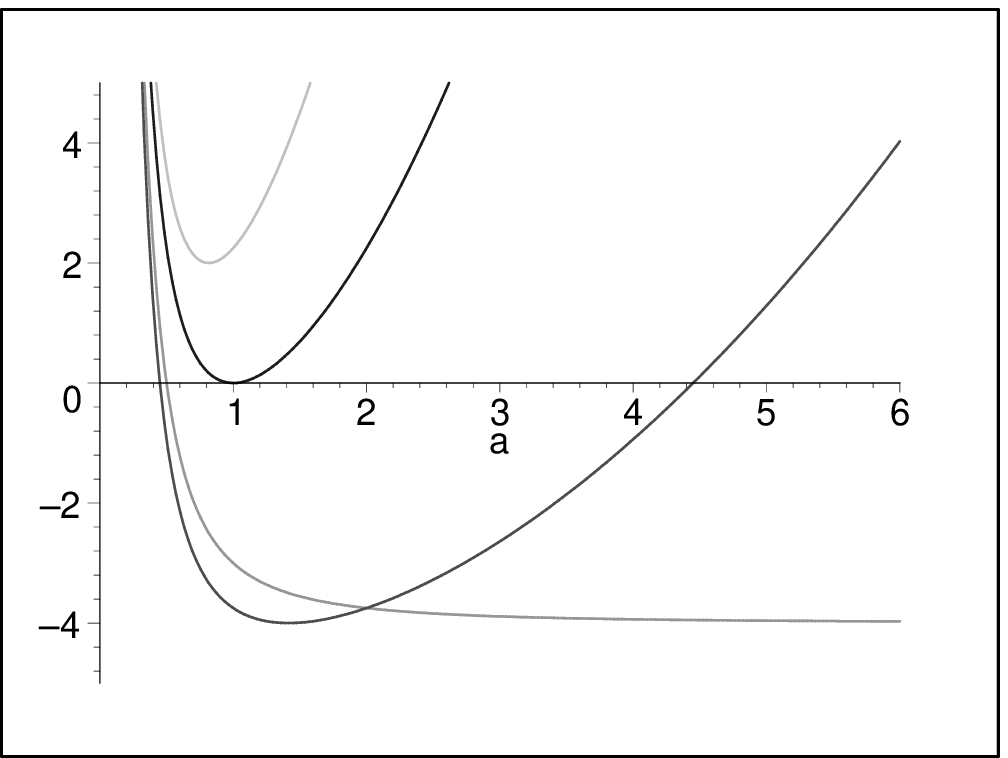}
%{Potential_k_negative_omega_positive01.eps}
\qquad\qquad
\includegraphics[width=7.3cm,keepaspectratio]{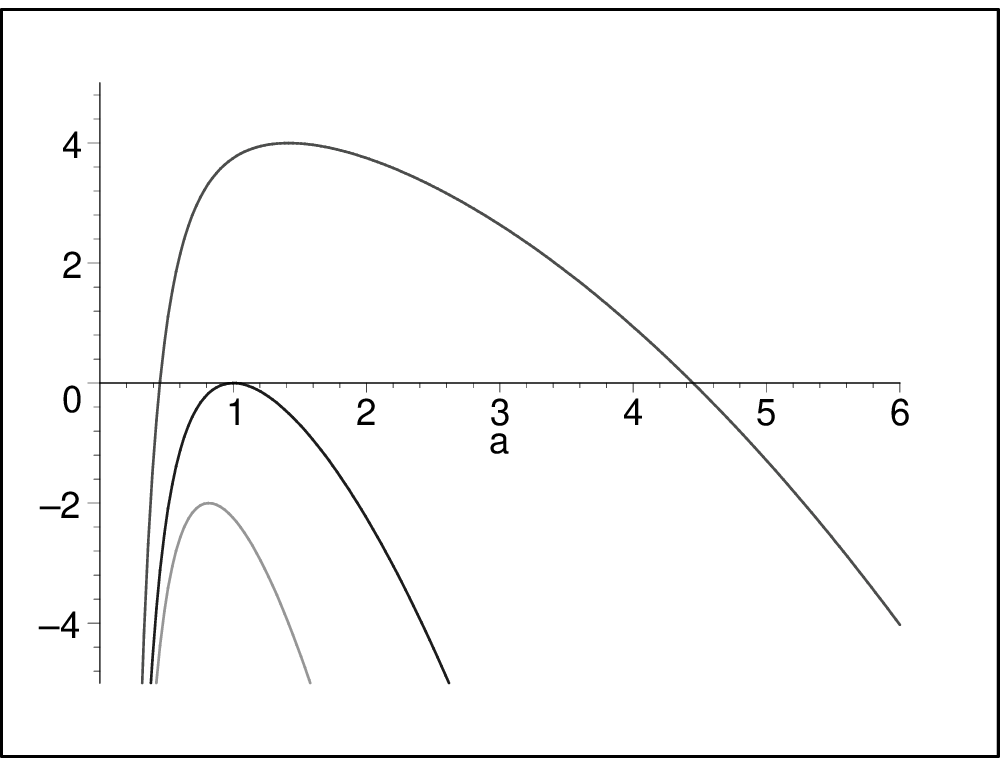}
%{Potential_k_positive_omega_negative01.eps}
\caption{Plots of the vacuum effective potential $V_{\rm eff}$ vs. scale
factor $a(t)$. \underline{Left}: Case $\om>0,\textbf{k}=-1$, in particular
$\La_W=-0.5,0,1,1.5, \om=2$ (bottom to top) with
$\kappa^4\mu^2/32(3 \lambda-1)^2\equiv 1$. The dynamical cosmology solutions
exist only for $\om (\om- 2 \La_W)>0, (\om-\La_W)>0$ and there are no initial
singularities at $a=0$. There are one bounce at $a^{-}$ and another at
$a^{+}$ for $\La_W\neq 0$ so that the universe becomes cyclic. But there is
only one bounce at $a^{-}$ for $\La_W= 0$ and so no cyclic universe exists.
\underline{Right}: Case $\om<0, \textbf{k}=+1$, in particular
$\La_W=0.5,-1,-1.5, \om=-2$ (top to bottom) with
$\kappa^4 \mu^2/32(3 \lambda-1)^2 \equiv -1$. The non-singular cosmology solutions, which have a bounce at $a^{+}$, exist only for $\om (\om- 2 \La_W)>0, (\om-\La_W)<0$ (top curve).} \label{fig:Potential_k_positive_omega_negative}
\end{figure}

More explicitly, for the AdS/flat branch with $\mu^2>0$, $\om>0$, in
particular, for $\La_W \neq 0$, the general solution for arbitrary {\bf k}
is given by
\begin{\eq}
a^2_{\rm AdS}(t)= \f{-{\bf k} (\om- \La_W)}{%R_0^2
\La^2_W} \left[ 1 +
\sqrt{\f{ \om (\om-2 \La_W)}{(\om -\La_W)^2}}~ \cos
\left(\f{\kappa^2 \mu \La_W}{%4
2 (3 \lambda-1)
} (t-\gamma) \right)\right]\, .
\label{AdS_Horava_FRW}
\end{\eq}
For $\om > 2 \La_W, {\bf k} (\om- \La_W)<0$, it admits a non-singular cyclic
cosmology solution, which is oscillating between the inner and outer bouncing
scale factors, $a^{-}$ and $a^{+}$, respectively, in (\ref{chonga}), with
an integration constant $\gamma$ depending on the initial conditions.
Notice that in this case, the second condition in (\ref{nonsingular_FRW})
can be satisfied only for $\mathbf{k}=-1$. In the case $\La_W=0$, on the
other hand, the general solution is given by
\begin{\eq}
a^2_{\rm flat}(t)=-\frac{\kappa^4 \mu^2 {\bf k} \om}{8 (3 \lambda-1)^2 } ~ (t-\gamma)^2
-\frac{{\bf k}}{2 \omega }
\label{Flat_Horava_FRW}
\end{\eq}
and, for ${\bf k}=-1$, this admits a non-singular cosmology solution with
only one bouncing at the scale factor $a^{-}$
(Fig. \ref{fig:Potential_k_positive_omega_negative} (left)).
Moreover, in the case $\om=0$, or $2 \La_W$, the two bouncing points meet and
there exists only a static cosmology solution of $a^2={\bf k}/\La_W$ or
-${\bf k}/\La_W$ when ${\bf k}=-1$ or $+1$, respectively, and $\La_W<0$ \ci{Lu}.

On the other hand, for the dS branch with $\mu^2<0, \om<0$, the general
solution for arbitrary {\bf k} is given by
\begin{\eq}
a^2_{\rm dS}(t)=\f{ 2 |3 \lambda-1| }{\kappa^2 |\mu| |\La_W|}
~{\rm e}^{\pm \f{\kappa^2 |\mu||\La_W|}{2 |3 \lambda-1|}
(t-\gamma)}+\f{{\bf k}^2 \kappa^2 |\mu| \om (\om-2 \La_W)}
{8 |3 \lambda-1| |\La_W|^3 }~ {\rm e}^{\mp \f{\kappa^2 |\mu| |\La_W|}
{2 |3 \lambda-1|} (t-\gamma)} -\f{ {\bf k}(\om- \La_W)}{\La_W^2}\, .
\label{dS_Horava_FRW}
\end{\eq}
For $\om < 2 \La_W, {\bf k} (\om-\La_W)<0$, the solution admits a
non-singular universe which has one bouncing at $a^+$ when $a(t)$ shrinks
toward $a^+$ from larger values and the universe evolves to $dS_4$ vacuum.
Notice that now the second condition in (\ref{nonsingular_FRW}) can be
satisfied only for $\mathbf{k}=+1$. Otherwise, the initial singularity
exists always. For example, for  $\om > 2 \La_W$ so that the first
condition in (\ref{nonsingular_FRW}) is not satisfied, there is a singular
solution with a bounce at $a^-$ when $a(t)$ expands toward $a^-$ from smaller
values and then shrink towards the initial singularity \footnote{For ${\bf k}\neq 0$, the solution (\ref{dS_Horava_FRW}) reduces to
$
%\begin{\eq}
a^2_{\rm dS}(t)= %\f
({-{\bf k} (\om- \La_W)}/{\La^2_W})
%\left
\{ 1 +\sqrt{
%\f
{ \om (\om-2 \La_W)}/{(\om -\La_W)^2}}~ \cosh
%\left
[%\f
({\kappa^2 |\mu||\La_W|}/{2 (3 \lambda-1)}) (t-\gamma)
%\right
]
%\right
\}\,
%\end{\eq}
$
or
$
%\begin{\eq}
a^2_{\rm dS}(t)= %\f
({-{\bf k} (\om- \La_W)}/{\La^2_W})
%\left
\{ 1 \pm \sqrt{
%\f
{ \om (-\om+2 \La_W)}/{(\om -\La_W)^2}}~ \sinh
%\left
[ %\f
({\kappa^2 |\mu||\La_W|}/{2 (3 \lambda-1)}) (t-\gamma)
%\right
]
%\right
\}\,
%\end{\eq}
$
by shifting the integration constant $\gamma$,
${\rm exp}[{\mp %\f
({\kappa^2 |\mu| |\La_W|}/{2 |3 \lambda-1|}) \gamma}]
\ra
%\f
\{{ \kappa^2 |\mu| |{\bf k}| \sqrt{\om (\om-2 \La_W)}}/
{ 4 |3 \lambda-1| |\La_W| }\}
~{\rm exp}[{\mp %\f
({\kappa^2 |\mu| |\La_W|}/{2 |3 \lambda-1|}) \gamma}]$
or
$
\{{ \kappa^2 |\mu| |{\bf k}| \sqrt{\om (-\om+2 \La_W)}}/
{ 4 |3 \lambda-1| |\La_W| }\}
~{\rm exp}[{\mp %\f
({\kappa^2 |\mu| |\La_W|}/{2 |3 \lambda-1|}) \gamma}]$
for $\om < 2 \La_W$ or $\om > 2 \La_W$, respectively. Here, the former non-singular case corresponds to the analytic continuation from the AdS branch solution (\ref{AdS_Horava_FRW}), but a similar continuation is absent for the latter singular case. For ${\bf k}=0$, on the other hand, (\ref{dS_Horava_FRW}) reduces to the exponentially growing or decaying solution also.}
(Fig. \ref{fig:Potential_k_positive_omega_negative} (right)).
Moreover, in the case $\om =0$ or $2 \La_W$, the two bouncing points meet
at $a^2= {\bf k}/\La_W$ or $-{\bf k}/\La_W$ but it admits the universe
which evolves monotonically from that minimum scale factor to $dS_4$ vacuum
asymptotically or vice versa \ci{Lu,Park:0905}.

Before concluding this section, we note that, as mentioned above, the
general (vacuum) cosmological solutions for arbitrary {\bf k} in GR can be
consistently obtained from the GR limit $(17)-(19)$ as follows :
\begin{\eq}
a_{\rm flat}(t)&=&\sqrt{-{\bf k} c^2}~  (t-\gamma)
~~~\mbox{for}~ \La=0, \no \\
a_{\rm AdS}(t)&=&\sqrt{-\f{3 {\bf k} c^2}{|\La|}}~ \cos \left(\sqrt{\f{|\La|}{3}} (t-\gamma)\right)
~~~\mbox{for}~ \La<0, \no \\
a_{\rm dS}(t)&=&\f{1}{\sqrt{12 \La}} \left(
\sqrt{6 \sqrt{3 \La}}~{\rm e}^{\pm \sqrt{
%\f
{\La}/{3}}(t-\gamma)}
+\f{9 {\bf k} c^2}{\sqrt{6 \sqrt{3 \La}}}
~{\rm e}^{\mp \sqrt{
%\f
{\La}/{3}}(t-\gamma)} \right)
~~~\mbox{for}~ \La>0.
\end{\eq}
In the last case, the solution reduces to the usual form of $a_{\rm dS}(t)=
\sqrt{3c^2/\La} \cosh (\sqrt{\La/3}(t-\gamma))$ or $a_{\rm dS}(t)=
\pm \sqrt{3c^2/\La} \sinh (\sqrt{\La/3}(t-\gamma))$ by the integration
constant shift of ${\rm e}^{\mp \sqrt{{\La}/{3}}\gamma}\ra (3 c /\sqrt{6 \sqrt{3 \La}})~{\rm e}^{\mp \sqrt{{\La}/{3}}\gamma}$ for ${\bf k}=+1$ or ${\bf k}=-1$, respectively. The recovery of cosmological solutions in GR, similarly to the black hole solutions in the previous sections, is not possible in the absence of $\om$, {\it i.e.}, in the original Ho\v{r}ava gravity with the detailed balance condition \ci{Lu}.

To conclude this section, we have classified the non-singular vacuum FLRW
cosmology solutions in Ho\v{r}ava gravity for the non-flat $({\bf k}\neq 0)$
universe, by the condition (\ref{nonsingular_FRW}). Note that the conditions
agree with the condition $\om (\om-2 \La_W)\geq0$ for the non-singular static
black hole/membrane geometry in Sec. II. And also from
(\ref{nonsingular_FRW}), we have found some intimate relation between
${\bf k}$ and $\om$, {\it i.e.,} ${\bf k}=-1$ for $\om>\La_W$, ${\bf k}=+1$
for $\om<\La_W$ for the non-singular cosmology solutions.

\section{Concluding Remarks }

We have studied the singularity and horizon structures of the
static black hole and membrane solutions in IR-modified Ho\v{r}ava
gravity, and classified all the viable solutions without naked singularities.
We have found a physical picture that is quite different from the
conventional one. In particular, we have found that, in addition to the usual
point-like singularity at the origin, there is a surface-like singularity
that becomes the cutting edge of space-time, where the real-valued metric
ends and unconventional complex-valued metric starts. The degrees of
divergence of curvature on such singularities is milder than that of GR.
Moreover, the Hawking temperature of the horizon is finite, unless any
of the singularities coincide with the the outermost horizon. We also found that there are viable black plane ($k=0$) and hyperbolic ($k=-1$) brane solutions even in the dS branch, where there is a cosmological horizon, as purely higher-derivative effects of Ho\v{r}ava gravity. We have also found some consistency with the conditions for non-singular time-dependent cosmological solutions. Several further remarks are in order.

First, according to Ho\v{r}ava gravity's idea for curing the renormalizability problem without ghosts, we need the higher-spatial derivative terms while
keeping quadratic in time-derivatives. However, wherever the lapse function
becomes negative, as it happens in the region inside the outer black hole
horizon or beyond the cosmological horizon, we have higher-time derivatives
while keeping quadratic in space-derivatives instead. But it is known that
the higher-time derivatives would produce the so called {\it Ostrogradsky
instability}. The detailed analysis would be beyond the scope of this paper, but we suspect that this may be not harmful at the classical level inside the
black hole/membrane horizons due to the finite range in the
``time"-coordinate $r$, that would prevent that a runaway behavior lasts
enough time to develop the infinite growth in any perturbations.
However, the problem persists beyond a cosmological horizon, if the real space-time does not end at some finite time $r_S$.

Second, the black plane solutions that we have studied can be also considered as the black ``string" solutions if we make a compactification along one direction on the plane \footnote{A certain class of the black string solutions, where the Cotton tensor vanishes, in the original Ho\v{r}ava gravity with the detailed balance condition are already known \ci{Cho:0909, Alie:1106}, but the general class of the black string solutions with/without the detailed balance condition are not known yet.}. It is well known that there is {\it Gregory-Laflamme instability} in higher-dimensional black strings for Einstein gravity. So, it would be interesting to investigate the similar instability in our four dimensional black plane solutions.

Third, the notion of horizon in Ho\v{r}ava gravity might be subtle, because it
depends on the dispersion relation of probing particles/fields. In
particular, if we consider the gravitational perturbations inside the
horizon, they can leak out from the horizon due to its non-relativistic
dispersions for high momentum, so that one can probe the singularities
inside the horizon. On the other hand, since the degree of singularity
is milder than that of GR, it would be interesting to investigate whether
it is possible to get some non-singular information via dispersive gravitons.

Fourth, we have found some interesting agreements in the conditions for non-singular static black hole or membrane metric with non-singular cosmology solutions. We do not know whether this is just a coincidence or there is some more fundamental reason which is not clear in our formulation.
Moreover, there is another type of correspondence which connects the domain-wall and cosmology solutions via complex coordinate transformations which mix space and time in GR \ci{Skenderis:2006fb} but does not hold anymore in Ho\v{r}ava gravity, due to lack of the symmetry between space and time. If there is some more fundamental reason for the obtained agreements, we may conjecture that a similar agreements may be found in this case also. It would be interesting to see whether there exits a similar correspondence between the non-projectable and projectable theories, which are known to be quite distinct in the original Ho\v{r}ava gravity setup that are adopted in this paper.

Finally, we have shown the importance of the ``GR limit", which achieves a
peculiar form of flows of coupling constants, in order to recover the results of GR in the asymptotic region. We do not have any fundamental understanding about such flow yet. In particular, it would be interesting to understand the relation of this flow and that of Renormalization Group.

\section*{Acknowledgments}

CRA and MIP are grateful to the organizers of the 13th Italian-Korean
meeting on Relativistic Astrophysics at Ewha Womans University,
%of Seoul
for providing a proper scientific environment which led to the original idea of this collaboration. NEG thanks Diana Lopez Nacir and Gast{\'o}n Giribet, and MIP would like to thank Shinsuke Kawai and Gungwon Kang for helpful comments. The authors are grateful to SISSA and ICTP for providing an extremely conformable work environment in which this work was started.

CRA was supported by the International Ceneter for Relativistic Astrophysics (ICRANet). NEG is supported by PIP CONICET grants N$^o$ 0396 and 0595 and UNLP project X648. MIP was supported by Basic Science Research Program through the National Research Foundation of Korea (NRF) funded by the Ministry of Education (2-2013-4569-001-1).

%%%%%%%%%% References %%%%%%%%%%%%%%%%%%%%%%%%%
\newcommand{\J}[4]{#1 {\bf #2} #3 (#4)}
\newcommand{\andJ}[3]{{\bf #1} (#2) #3}
\newcommand{\AP}{Ann. Phys. (N.Y.)}
\newcommand{\MPL}{Mod. Phys. Lett.}
\newcommand{\NP}{Nucl. Phys.}
\newcommand{\PL}{Phys. Lett.}
\newcommand{\PR}{Phys. Rev. D}
\newcommand{\PRL}{Phys. Rev. Lett.}
\newcommand{\PTP}{Prog. Theor. Phys.}
\newcommand{\hep}[1]{ hep-th/{#1}}
\newcommand{\hepp}[1]{ hep-ph/{#1}}
\newcommand{\hepg}[1]{ gr-qc/{#1}}
\newcommand{\bi}{ \bibitem}
%%%%%%%%%%%%%%%%%%%%%%%%%%%%%%%%%%%%%%%%%%%%%%%

\end{document}